\documentclass[a4paper,12pt]{article}
\def\al{\alpha}
\def\be{\beta}
\def\ga{\gamma} \def\Ga{\Gamma}
\def\ep{\epsilon}
\def\lam{\lambda}
\def\Lam{\Lambda}

  \def\calO{{\cal O}}
 \def\calQ{{\cal Q}}

\def\shead#1{\parbigskipn {\bf #1} \parmedskipn} 
\def\del        {  \partial  }
\def\half       {  {1\over 2}  }
\def\trace      {  \mbox{Tr}  }

\def\ie         {  {\it i.e.}      }

\def\comma          {\, ,}
\def\period         {\, .}
\def\lsim    {\lower .65ex \hbox{\ $\stackrel{<}{\sim}$\ } }
\def\gsim    {\lower .65ex \hbox{\ $\stackrel{>}{\sim}$\ } }
\def\com#1#2   { \left[#1, #2\right]} 
\def\acom#1#2  {\left\{ #1,#2\right\}}
\def\bra#1     {\langle #1 |}
\def\ket#1     {| #1 \rangle}
\def\vecii#1#2      {  \left(\begin{array}{c}#1\\#2\end{array}\right)  }
\def\veciii#1#2#3   {  \left(\begin{array}{c}#1\\#2\\#3\end{array}
                     \right)  }
\def\veciv#1#2#3#4  {  \left(\begin{array}{c}#1\\#2\\#3\\#4
                                 \end{array}\right)  }
\def\vecfv#1#2#3#4#5 {  \left(\begin{array}{c}#1\\#2\\#3\\#4\\#5
                                 \end{array}\right)  }

\def\matrixii#1#2#3#4            {  \left(\begin{array}{cc}#1&#2\\#3&#4
                                       \end{array}\right) }
\def\matrixiii#1#2#3#4#5#6#7#8#9 {  \left(\begin{array}{ccc}#1&#2&#3\\
                                     #4&#5&#6\\#7&#8&#9\end{array}
                               \right)  }
\def\mativ#1#2#3#4               {  \left(\begin{array}{cccc}
                                       #1\\#2\\#3\\#4\end{array}\right) }
\def\matv#1#2#3#4#5              {  \left(\begin{array}{ccccc}
                                     #1\\#2\\#3\\#4\\#5\end{array}
                              \right)  }
\def\eqabegin         {  \begin{eqnarray}  }
\def\eqaend           {  \end{eqnarray}  }
\def\nn               {  \nonumber  }
\def\bracetwo#1#2     {  \left\{ \begin{array}{l} #1 \\ #2 \end{array}
                         \right.  }
\def\bracetwocases#1#2#3#4  {   \left\{ \begin{array}{ll} #1 &
                                 \qquad #2 \\
                                 #3 & \qquad #4 \end{array} \right.  }
\def\bracebegin#1     {  \left\{ \begin{array}{#1}   }
\def\braceend         {  \end{array}\right.   }
\def\parn              {  \par\noindent }

\def\parmedskip        {  \par\medskip  }
\def\parsmallskip      {  \par\smallskip  }
\def\parbigskipn        {  \par\bigskip\noindent  }
\def\parmedskipn        {  \par\medskip\noindent  }

\def\parag#1           {\paragraph{#1} \mbox{ }\parmedskip\noindent}
\def\msection#1      {  \begin{center} \section{#1} \end{center}   }
\def\nsection#1      {  \let\boldface\bf \def\bf{} \section{#1}
                           \let\bf\boldface   }
\def\mnsection#1     {  \begin{center} \nsection{#1} \end{center}  }
\def\capsection#1    {  \let\boldface\bf \def\bf{\sc} \section{#1}
                           \let\bf\boldface   }
\def\mcapsection#1   {  \begin{center} \capsection{#1} \end{center} }

\def\sectionnumbering { \setcounter{equation}{0}
         \renewcommand{\theequation}{\arabic{section}.\arabic{equation}}}

\newcommand{\nullify}[1]{}
\def\papertitlepage{\baselineskip 3.5ex \thispagestyle{empty}}
\def\Title#1{\baselineskip 1cm \vspace{1.5cm}\begin{center}
 {\Large\bf #1} \end{center} 
\vspace{0.5cm}}
\def\Authors#1{\begin{center} {\it #1} \end{center}}
\def\Abstract{\vspace{1.0cm}\begin{center} {\large\bf Abstract} 
           \end{center} \par\bigskip}
\def\Komabanumber#1#2#3{\hfill \begin{minipage}{4.2cm} UT-Komaba #1
              \parn #2 
              \parn #3 \end{minipage}}
\renewcommand{\thefootnote}{\fnsymbol{footnote}}
\renewenvironment{thebibliography}{\pagebreak[3]\par\vspace{0.6em}
\begin{flushleft}{\large \bf References}\end{flushleft}
\vspace{-1.0em}

\begin{enumerate}\if@twocolumn\baselineskip=0.6em\itemsep -0.2em
\else\itemsep -0.2em\fi\labelsep 0.1em}{\end{enumerate}}

\def\Gatil{\widetilde{\Ga}}

\def\Cbar{\overline{C}}

\def\Atil{\widetilde{A}}

\def\thdot{\dot{\theta}}

\def\Stil{\widetilde{S}}

\def\Gtil{\widetilde{G}}

\def\yhat{\widehat{y}}
\def\psihat{\widehat{\psi}}
\def\slash#1{#1\!\!\!/}

\def\tilA{\widetilde{A}}
\def\lk{\langle}
\def\rk{\rangle}
\def\taup{\tau'}
\def\taupp{\tau''}
\def\shead#1{\parmedskipn {\large\bf #1}}
\newcommand{\barr}{\begin{array}}
\newcommand{\earr}{\end{array}}
\def\tilA{\widetilde{A}}
\def\tilS{\widetilde{S}}
\def\dirac#1{{\ooalign{\hfil/\hfil\crcr$#1$}}}
\newcommand{\p}{\partial}
\def\brkeq{\nonumber \\[1.2ex] &}
\def\brk{\nonumber \\[0.5ex] &}
\def\and{&}
\usepackage{graphicx}
\usepackage{latexsym}
\usepackage{amsmath}
\usepackage{amssymb}
\def\ind#1#2{#1_1\ldots #1_#2}
\def\indr#1#2{#1_#2\ldots #1_1}
\renewenvironment{thebibliography}{\pagebreak[3]\par\vspace{0.6em}
\begin{flushleft}{\large \bf References}\end{flushleft}
\vspace{-1.0em}

\begin{enumerate}\if@twocolumn\baselineskip=0.6em\itemsep -0.2em
\else\itemsep -0.2em\fi\labelsep 0.1em}{\end{enumerate} }
\begin{document}
\papertitlepage
\vspace*{0cm}
\Komabanumber{01-01}{hep-th/0103116}{March, 2001}
\Title{Fully Off-shell Effective Action and its Supersymmetry in
 Matrix Theory} 
\vspace{1cm}
\Authors{{\sc Y.~Kazama\footnote[2]{kazama@hep3.c.u-tokyo.ac.jp} 
 and T.~Muramatsu
\footnote[3]{tetsu@hep1.c.u-tokyo.ac.jp}
\\ }
\vskip 3ex
 Institute of Physics, University of Tokyo, \\
 Komaba, Meguro-ku, Tokyo 153-8902 Japan \\
  }
\baselineskip .7cm
\Abstract
As a step toward clarification of the power of supersymmetry (SUSY)
 in Matrix theory,
 a complete calculation, including all the spin effects,
 is performed of the effective action of a probe
 D-particle, moving along an {\it arbitrary} trajectory
 in interaction with a large number of coincident source D-particles, 
at one loop at order 4 in the derivative expansion. Furthermore, 
exploiting the SUSY Ward identity developed previously, 
the quantum-corrected effective supersymmetry transformation laws 
are obtained explicitly to the relevant order and 
are used to verify the SUSY-invariance of the effective action. 
Assuming that the agreement with 11-dimensional supergravity persists, 
 our result can be regarded as a prediction for supergravity 
calculation, which, yet unavailable,  is known to be highly non-trivial. 
 
\newpage
\baselineskip 3.5ex
\section{Introduction}  
 \sectionnumbering
\renewcommand{\thefootnote}{\arabic{footnote}}
One of the main themes of string theory in the past few years is the 
 correspondence between supergravity/superstring theories in the bulk 
and certain types of  supersymmetric theories defined on the worldvolume 
 of various brane configurations or \lq\lq on the boundary" of the 
 space-time. The proposal of Maldacena\cite{maldacena}, termed AdS/CFT 
 correspondence, has by now been extended to much wider class of systems
 than was originally envisaged and has spurred wide variety of
 new developments\cite{adscft}. 
\parsmallskip
Although conjectured prior to this proposal, the 
Matrix theory for M-theory, put forward by Banks, Fischler, Shenker 
and Susskind \cite{bfss}, can be thought of as a prime example of this
 correspondence. Re-interpreted in the frame work of 
discrete light-cone quantization \cite{susskind}, 
it has enjoyed numerous successes. 
Just to mention only the direct comparison 
with eleven dimensional 
 supergravity,  complete agreement for the multi-graviton scattering
 (including the recoil effects) at 2-loop\cite{Okawa-Yoneya,Okawa-Yoneya2}
 and 
 that for the two-body potential between arbitrary fermionic as well as 
 bosonic objects at 1-loop
\cite{TR} should be regarded as highly non-trivial and remarkable. 
\parsmallskip
Just as the mechanism of the general bulk-boundary correspondence 
 has not been clearly identified, the origin of these
 successes in the Matrix theory is yet to be fully understood. 
Now in string theory, it is often the case that 
symmetry principles play decisive roles, much more so than in 
 local field theories, in determining the dynamics of the system, 
 and the situation should be the same in Matrix theory, which is
 an explicit representation of M-theory that unifies all the string
 theories. 
\parsmallskip
 Indeed there have been a number of studies\cite{Pabanetal1, Pabanetal2, 
 lowe, Hyunetal, ss, np} which 
 point to the assertion that, in particular, 
 the high degree of supersymmetry, 
 namely with the maximally allowed $16$ supercharges, is powerful enough 
 to determine the effective action of the D-particles, which can be
 directly compared with the corresponding supergravity calculation. 
If it is indeed the case, it is rather surprising since usually 
 such a global symmetry can only give certain relations among 
 the correlation functions and not more,  and consequently 
 the dynamical significance of Matrix theory would have to be 
 reconsidered. However, upon close examinations, even restricted to 
 the simple so-called \lq\lq source-probe" situation,
 one can argue that the existing analysis is not quite complete.
  This is largely due to the fact that for a system with maximal supersymmetry
 unconstrained superfield formulation does not exist and therefore
 a clear-cut off-shell analysis is not possible: One is forced to 
 deal with the component formalism, where the supersymmetry algebra 
 gets intertwined with gauge symmetry and does not close without the 
 aid of equations of motion. 
\parsmallskip
In a previous work\cite{Kaz-Mura}, we gave a rather extensive discussion on 
 the existing literature and emphasized the importance of the 
 consistency of the approximation scheme in studying the symmetry of the 
effective action, which is an off-shell entity. We argued  that 
 the only consistent procedure  is to deal with the 
 trajectory, including the spin degrees of freedom, with  {\it arbitrary} time 
 dependence and adopt the derivative expansion
 according to the concept of {\it order}, defined as the number of derivatives 
 plus  half the number of fermions.  Only this scheme is free of 
 total derivative ambiguities inevitably present in the effective action. 
For more discussions, see the Section 3 of \cite{Kaz-Mura}. 
\parsmallskip
Following this philosophy, we derived in \cite{Kaz-Mura} 
 a completely off-shell SUSY Ward identity in the background gauge
 (which is naturally intertwined with the BRST symmetry) and applied it
 to the effective action at order 2 to study the power of supersymmetry.
Our conclusion was that at this order the effective action is 
 indeed determined by the Ward identity alone. 
As we already remarked there, however, such a result was to 
 be expected since at order 2 the higher derivatives, such as the 
 acceleration etc., can be eliminated from the effective action
 by integration by parts and our analysis was essentially the same as in 
 the existing literature. The full significance of the completely off-shell
 analysis becomes apparent starting from the next order, \ie from order 4,
 where complete elimination of higher derivatives is no longer possible. 
\parsmallskip
 Unlike the case of order 2, at which the 1-loop 
effective action was already available,  the only fully off-shell results
 so far known at order 4 
are the famous bosonic part\cite{dkps, hm9901, okawa9903} given (in Euclidean 
 formulation) by $-N\int d\tau (15/16) v^4/r^7$, with $N$  the number
 of source D-particles,  and a part of the fermionic 
 contributions
 containing 8 powers of the fermion field $\theta$ \cite{Barrioetal}.
 The latter is the simplest among the fermionic contributions 
 since it does not involve any derivatives. 
 A calculation of $\calO(\theta^2)$ contribution, which is already 
 formidable,  was attempted in 
 \cite{okawa9907} but was not fully completed. As for $\calO(\theta^4)$
and $\calO(\theta^6)$ contributions, no attempt has been made to date.  
\parsmallskip
In this work, as a necessary step toward clarification of 
the power of supersymmetry in Matrix theory, we perform a complete 
 calculation of the effective action for a probe D-particle, including
 all the spin effects,  at one-loop at order 4, with the aid of the 
 algebraic manipulation program Mathematica. Furthermore, 
exploiting the SUSY Ward identity developed previously, 
the quantum-corrected effective supersymmetry transformation laws 
are obtained explicitly to the relevant order and 
are used to verify the SUSY-invariance of the effective action so obtained.
Since the corresponding supergravity calculation is available only 
 up to $\calO(\theta^2)$ \cite{HyunetalPRD},
 we cannot at present time test if our result agrees
 with supergravity. Rather, provided that the agreement with 
 supergravity persists, our result should be regarded as a prediction 
until such a calculation will have been made. 
\parsmallskip
Organization of the rest of this article is as follows: We start in
Section 2  with 
 a brief summary of the Matrix theory and its symmetries, mainly to set
 the notations. Section 3 is devoted to the calculation of the 
 effective action at one-loop. After explaining the derivative expansion
 scheme used and reviewing the existing results for the effective action
 in Section 3.1, we describe the calculational procedures in Section
 3.2. The result of the calculation, together with how it was simplified using 
 $SO(9)$ Fierz identities, is presented in Section 3.3. Since the actual
 calculation, performed by Mathematica, involved an enormous number of steps,
 it is desirable to have an independent check. For this purpose, 
 as well as for its own interest, we compute in Section 4 the 
quantum-corrected SUSY transformation laws
and check the invariance of the effective action under these transformations:
Following a brief review of the SUSY  
Ward identity in Section 4.1, 
 we sketch the calculational procedure in Section 4.2. Then 
 in Section 4.3 we describe how the invariance under SUSY was verified. 
Finally, we give a summary and discussions in Section 5. 
Three appendices, A$\sim$C are provided to supply  some details of 
 the calculations. In appendix A we describe a new efficient algorithm 
 for generating $SO(9)$ Fierz identities. Non-trivial two-point functions
 needed for the calculation of the SUSY transformation laws are 
 collected in Appendix B and  the quantum-corrected SUSY 
 transformation laws are displayed in Appendix C. 
\section{Preliminaries}
We begin with a very brief summary of the Matrix theory and its symmetries,
 mainly to set our notations. 
\parsmallskip
The classical action for the $U(N+1)$
 Matrix theory in the Euclidean formulation 
 is given by 
\begin{align}
\Stil_0 &= \trace \int\!\!d\tau  \ \Biggl\{ \frac{1}{2} \com{D_\tau}{X_m} ^2 
- \frac{g^2}{4}
 \com{X_m}{X_n} ^2  \nn\\
& \qquad +\frac{1}{2}\Theta^T \com{D_\tau}{\Theta} 
-\frac{1}{2} g \Theta^T \ga^m\com{X_m}{\Theta} \Biggr\}\period
\end{align}
In this expression, 
$X^m_{ij}(\tau), \Atil_{ij}(\tau)$ and $\Theta_{\al,ij}(\tau)$ are 
the $(N+1)\times (N+1)$ 
hermitian matrix fields, representing the bosonic part of the
D-particles, the gauge fields, and the fermionic part of the D-particles,
 respectively.  $D_\tau = \del_\tau -ig\Atil$ is the covariant derivative,
 $\ga^m$ are the real symmetric $16\times 16$ $SO(9)$ $\ga$-matrices, and 
the vector index $m$ runs from $1$ to $9$. We put a tilde on some relevant 
symbols to remind us of Euclidean formulation. 
\parsmallskip
This action is known to possess a number of important symmetries. The first 
 is the obvious global Spin(9) invariance, which nevertheless is quite 
 non-trivial in the fermionic sector being 
responsible for various Fierz identities to be used extensively later. 
The second is the CPT symmetry inherited from the 10-dimensional 
 super Yang-Mills theory from which the above action can be obtained by 
 dimensional reduction. The third is the invariance under the 
$U(N+1)$ gauge transformations given, with the gauge parameter matrix $\Lam$,
 by
\begin{align}
\delta_\Lam \Atil &= \com{D_\tau}{\Lam} \comma \qquad 
\delta_\Lam X_m = ig\com{\Lam}{X_m} \comma  \nn\\
\delta_\Lam \Theta &= ig\com{\Lam}{\Theta} \period 
\end{align}
The fourth, and the main focus of this article,
 is the supersymmetry with 16 spinorial parameters $\ep_\al$.
The transformation laws are 
\begin{align}
\delta_\ep \Atil &= \ep^T \Theta \comma \qquad 
\delta_\ep X^m = -i\ep^T \ga^m\Theta \comma \label{treesusy}\\
\delta_\ep \Theta &= i\left( \com{D_\tau}{X_m} \ga^m
 +{g\over 2} \com{X_m}{X_n} \ga^{mn} \right) \ep \period
\end{align}
Although the algebra closes only on-shell up to field-dependent 
 gauge transformations, $\Stil_0$ itself is invariant without the use of 
equations of motion, \ie off-shell. 
\parsmallskip
In addition to these well-known symmetries, there is a so-called 
 generalized conformal symmetry\cite{jy,jky,jky2}, which may be used to 
 restrict the form of the effective action. Finally, the agreement 
with the 11-dimensional supergravity calculation for the multi-body
 processes\cite{Okawa-Yoneya} strongly suggests that the 11-dimensional 
 Lorentz symmetry is hidden in $\Stil_0$, awaiting to be disclosed.
\parsmallskip
In this article, we shall concentrate on the so-called source-probe 
situation, namely the configuration of a probe D-particle 
interacting with a large number, $N$, of the source D-particles all sitting 
 at the origin. This is expressed by the splitting 
\begin{alignat}{2}
X_m(\tau) &= {1\over g} B_m(\tau) +Y_m(\tau) \comma &  
\qquad \Theta_\al(\tau) &= {1\over g} \theta_\al(\tau)
+ \Psi_\al(\tau)\comma  \\
B_m(\tau) &= {\rm diag}\, (r_m(\tau), \overbrace{0,0,\ldots , 0}^N) \comma &
\qquad \theta_\al(\tau) &=  {\rm diag}\, 
 (\theta_\al(\tau), \overbrace{0,0,\ldots , 0}^N) \comma 
\end{alignat}
where $B_m(\tau)$ and $\theta_\al(\tau)$ are 
 the bosonic and the fermionic backgrounds expressing the 
positions and the spin degrees of freedom of the D-particles 
 respectively 
 and $Y_m(\tau)$ and $\Psi_\al(\tau)$ denote the quantum 
fluctuations around them. We will be interested in the general case
 where $r_m(\tau)$ and $\theta_\al(\tau)$ are {\it arbitrary functions
 of $\tau$} not satisfying equations of motion. 
\parsmallskip
In order to quantize the system and perform the calculation of the effective
 action of the probe D-particle, we must fix the gauge. Practically 
 the only tractable choice, and indeed the one used for all the 
 calculations in the past, is the background gauge specified by 
 the gauge-fixing function $\Gtil$ given by 
\begin{equation}
\Gtil = -\del_\tau \Atil + i\com{B^m}{X_m} \label{sbgg} \period
\end{equation}
The associated BRST transformations for the quantum part of the 
fields are given by
\begin{align}
\delta_{\rm B} \Atil &=  \com{D_\tau}{C} \comma \qquad 
\delta_{\rm B} Y_m = -ig\com{X_m}{C} \comma \nn\\
\delta_{\rm B} \Psi &= ig\acom{C}{\Theta} \comma \label{brstr}\\
\delta_{\rm B} C &= igC^2 \comma \qquad \delta_{\rm B} \Cbar = ib
\comma \qquad \delta_{\rm B} b=0\comma  \nn
\end{align}
where $C, \Cbar$ and $b$ are, respectively, the ghost, the anti-ghost
 and the Nakanishi-Lautrup auxiliary field. As usual, 
the gauge-fixing and the ghost part of the action are given altogether
 by the BRST total variation
\begin{equation}
\Stil_{\rm gg} = \delta_{\rm B} \trace \!\int\!\! d\tau \left[ {1\over i} \Cbar \left(\Gtil-
{1 \over 2} b\right) \right]  \label{sgg} \period 
\end{equation}
\section{Calculation of the Off-shell Effective Action}
We are now ready to start the computation of the off-shell effective
 action for the probe D-particle. 
\subsection{Derivative expansion and the existing results}
As was already emphasized in the introduction and further elaborated 
 in \cite{Kaz-Mura}, it is important that the approximation scheme 
for computing the effective action 
 must be consistent with the freedom of adding total derivatives. 
The only such scheme is the derivative expansion according to the 
 concept of {\it order} defined as \cite{harvey} 
\begin{equation}
{\rm order} \equiv  \mbox{number of $\tau$-derivatives} + \half \mbox{number
 of fermions} \period
\end{equation}
In other words, we assign ${\rm order} (r)=0\comma {\rm order}(\del) = 1$,
 and ${\rm order}(\theta)=1/2$, where $\del$ denotes the derivative with 
 respect to $\tau$. 
\parsmallskip
This concept can be applied loop-wise. For instance, the
 tree level action for $r_m$
 and $\theta_\al$ is of the form (we use $v_m$ to denote $\dot{r}_m$,
the superscript in parenthesis is the loop number and the subscript
 signifies the order)
\begin{equation}
\Gatil^{(0)}_2 = \int\!\! d\tau \left( {v^2\over 2g^2}
 + {\theta^T \thdot\over 2g^2}\right) \comma 
\end{equation}
and is entirely of order 2. The order 2 part at 1 loop was first 
 computed\footnote{Here and hereafter, all the results are
 in the background gauge.} in \cite{TR} and is given by
\begin{equation}
\Gatil_2^{(1)} = 
 N\! \int\!\! d\tau  \ {\theta^T\thdot \over r^3} \period
\end{equation}

Consider now the order 4 part at 1-loop, which consists of the bosonic
 and the fermionic parts. 
The former was computed long ago in eikonal approximation in \cite{dkps}
and more recently for fully off-shell case in \cite{hm9901, okawa9903}
and has the well-known form
\begin{equation}
\Gatil_1^{(4),b} = -N\!\int\!\! d\tau  \ {15 \over 16}{v^4\over r^7}\period
\label{dkps}
\end{equation}
As we shall see shortly, calculation of the fermionic part
\footnote{Within the eikonal-type approximation, spin effects have been
discussed in\cite{harvey, moralesetal97, kraus, moralesetal98, 
plefkaetal9806, mcarthur, plefkaetal9809}.}%
, which
 can be further classified by the (even) number of $\theta$'s up to 8, 
 is exceedingly
 more difficult. The $\calO(\theta^8)$ part is relatively easy to compute
\cite{Barrioetal} since it cannot contain any derivatives. 
The calculation of the $\calO(\theta^2)$ part (with three $\del$'s) 
was attempted in 
\cite{okawa9907} but was not fully completed. As for the $\calO(\theta^4)$
 and $\calO(\theta^6)$ contributions, even an attempt has not been made. 
\parsmallskip
In the following, we perform the complete calculation of all 
the fermionic order 4 terms at 1 loop. This is made possible with 
the extensive use of Mathematica. 
\subsection{Calculational procedure}
As we shall perform the calculation at 1 loop, we only need the part 
or the gauge-fixed action quadratic in the quantum fluctuations. To simplify
 the presentation as well as the subsequent calculations, define the 
 $(1+9)$-component fermionic and bosonic vectors in the following way:
\begin{eqnarray}
\Xi_\alpha \equiv \vecii{i\theta_\al}{\theta_\be \ga^m_{\be\al}} 
\comma \quad \Phi \equiv \vecii{\!\!\!\!\Atil}{Y^m} \comma \label{XiPhi}
\end{eqnarray}
where the matrix indices are suppressed.  Also define the 
 kinetic operators $D_{\rm B}$ and $D_{\rm F}$ 
for bosonic and fermionic fields respectively as
\begin{equation}
D_{\rm B} \equiv  \matrixii{\Delta^{-1}}{-2iv^m}{2iv^n}{\Delta^{-1}\delta_{nm} } 
\comma\qquad  D_{\rm F} \equiv \del + \slash{r} \period \label{DBDF}
\end{equation}
Here, $\Delta^{-1}\equiv -\del^2+r^2(\tau)$ 
is the basic kinetic operator, and its inverse $\Delta$,
the basic propagator, will appear frequently in the actual calculations. 
Note that the \lq\lq mass" $r(\tau)$ is an arbitrary function of $\tau$
 and this will make the computation non-trivial. 
\parsmallskip
With these notations, the quadratic part of the action can be written as
\begin{align}
\tilS_{(2)}&=\int\!\! d\tau \sum_{i,j}\Bigg[
\frac{1}{2}\Phi_{ij} D_{\rm B} \Phi_{ji}
+ i\Cbar_{ij} \Delta^{-1} C_{ji} \nn \\
& \hspace{1cm} + \frac{1}{2}\Psi_{ij} D_{\rm F}\Psi_{ji} +
\frac{1}{2}\Phi_{ji}^T\Xi \Psi_{ij}
+ \frac{1}{2} \Psi_{ij}\Xi^T \Phi_{ij}\Bigg] \period \label{quadaction}
\end{align}
Performing the Gaussian integration,  we obtain the formal expression
 for the 1-loop effective action $\widetilde{\Gamma}^{(1)}$ consisting 
 of the bosonic and the fermionic parts:
\begin{align}
\widetilde{\Gamma}^{(1)} &=  \Gatil^{(1)}_{\rm B}+\Gatil^{(1)}_{\rm F} \comma  \\
\Gatil^{(1)}_{\rm B} &= 
{\rm tr} \ln ( 1 - 4 v^n \Delta v^n \Delta )
- \frac{1}{2} {\rm Tr} \ln
( {\bf 1} + \dirac{v} \Delta )\comma \nn\\
\Gatil^{(1)}_{\rm F}&=
-{\rm Tr}\ln \Big( {\bf 1} - D_{F}^{-1}\Xi^{T} D_{\rm B}^{-1}\Xi\Big)
=\sum^{\infty}_{n=1}
\frac{
{\rm Tr}\left(D_{\rm F}^{-1}\Xi^T D_{\rm B}^{-1}\Xi\right)^n
}{n} \period
\end{align}
In the above, \lq\lq tr" is the trace over the matrix indices and 
 over the function space, while \lq\lq Tr" further includes the trace
 over the spinor indices as well. At order 4, the bosonic part $\Gatil^{(1)}_{\rm B}$
reproduces the known result (\ref{dkps}). Our interest is in the fermionic
 part $\Gatil^{(1)}_{\rm F}$ and we need to compute up to $n=4$. 

Now to proceed, we must expand the propagators $D_{\rm B}^{-1}$ and $D_{\rm F}^{-1}$
 in powers of the derivatives. The explicit form of $D_{\rm B}^{-1}$ is given 
by 
\begin{align}
D_{\rm B}^{-1}&= 
\left( 
\barr{cc}
\Delta (1 - 4v^\ell\Delta v^\ell\Delta)^{-1} &
 2i\Delta (1 - 4v^\ell\Delta v^\ell\Delta)^{-1} v^m \Delta \\
-2i\Delta v^n \Delta(1 - 4v^\ell\Delta v^\ell\Delta)^{-1} &
 \Delta (\delta_{nm} -4v^n\Delta v^m\Delta )^{-1}
\earr
\right) \comma \label{Bprop}
\end{align}
where the expansion of the following type should be substituted:
\begin{align} 
 (\delta_{nm} -4v^n\Delta v^m\Delta )^{-1} \equiv 
\delta_{nm} &+ 4v^n\Delta v^m\Delta +
16 v^n\Delta v^\ell\Delta v^\ell\Delta v^m\Delta +\cdots \period
\end{align}
The corresponding expansion for $D_{\rm F}$ is given by 
\begin{align}
 D_{\rm F}^{-1} &= (\p + \dirac{r})^{-1}=
-(\p - \dirac{r})
(1 + \Delta\dirac{v})^{-1}\Delta \nn\\
&= -(\p - \dirac{r})\Delta+(\p-\dirac{r})\Delta \dirac{v} \Delta +
\cdots \period \label{Fprop}
\end{align}
When these expansions are implemented, evidently the expressions for ${\rm Tr}
\left(D_{\rm F}^{-1}\Xi^T D_{\rm B}^{-1}\Xi\right)^n$ get rather involved. 
Below we only display the intermediate result (which 
still has to be fully expanded) for the simplest case of 
$n=1$ as an example:
\begin{align}
&{\rm Tr}D_{\rm F}^{-1}\Xi^T D_{\rm B}^{-1}\Xi 
\nn \\ 
&\quad ={\rm Tr} \Big[
-(\p - \dirac{r})
(1 - \Delta\dirac{v} + \Delta\dirac{v}\Delta\dirac{v}
 -\Delta\dirac{v}\Delta\dirac{v}\Delta\dirac{v}
)_{\alpha\epsilon}\Delta \nn \\
& \qquad\ \times 
\Big(
-\theta_\epsilon \Delta \theta_\alpha  
-4\theta_\epsilon \Delta v^\ell \Delta v^\ell \Delta \theta_\alpha 
-2\Delta\theta_\epsilon \dirac{v}_{\delta\alpha}\Delta\theta_\delta 
-8 \Delta\theta_\epsilon v^\ell\Delta v^\ell\Delta 
\dirac{v}_{\delta\alpha}  \Delta\theta_\delta 
\nn \\
& \qquad\ 
+
2\theta_\gamma \Delta \dirac{v}_{\gamma\epsilon} \Delta \theta_\alpha  
+ 8 \theta_\gamma \Delta \dirac{v}_{\gamma\epsilon}
\Delta v^\ell\Delta v^\ell\Delta \theta_\alpha 
+\theta_\gamma \gamma_{\gamma\epsilon}^n \Delta  
\theta_\delta \gamma^n_{\delta\alpha}  \nn\\
&\qquad\ + 4 \theta_\gamma \Delta 
\dirac{v}_{\gamma\epsilon}\Delta \dirac{v}_{\delta\alpha}\Delta \theta_\delta 
\Big)  \Big]
+ \calO (\p^4) \period
\end{align}
The number of terms are much larger for $n\ge 2$ and altogether 
 they add up to about 1000. 
\parsmallskip
The next step is the actual evaluation of the trace \lq\lq Tr", which here
means tracing over the function space and over the spinor indices. 
While the latter, which is nothing but the $\ga$-matrix algebra,
 is conceptually simple, the former step requires some explanation. 
To perform it efficiently, we employ the so-called 
\lq\lq normal-ordering method", an algorithm invented by Okawa in 
\cite{okawa9903}. The idea is to bring the expressions involving 
 $f(\tau)$ (some function of $\tau$), $\del$ and $\Delta$ in various orders
 into a \lq\lq normal-ordered" form $f(\tau) \del^m \Delta^n$
 by recursively using the formulas
\begin{align}
\del f &= f\del + \dot{f} \comma \nn\\
\Delta f &= f \Delta +\Delta(\ddot{f} + 2\dot{f} \del) \Delta \comma \\
\Delta \del &= \del \Delta + 2\Delta (\dot{r}\cdot r) \Delta \period \nn
\end{align}
Further, by using the trivial relation  $\del^2\Delta=r^2\Delta -1$, 
which follows from the definition of $\Delta$, 
 one can reduce the number of $\del$'s in the normal-ordered form down to 
 either zero or one. 
Once each term is brought to this standard form, the remaining task is to 
 compute the matrix elements $\langle  \tau| \Delta^n | \tau'\rangle$ 
 and $\langle  \tau| \del \Delta^n | \tau'\rangle$, where the latter is 
 actually expressed in terms of the former as 
$\langle  \tau| \del \Delta^n | \tau'\rangle=\half \del 
\langle  \tau| \Delta^n | \tau'\rangle$. Now employing an integral 
 representation, one can write
\begin{align}
\langle  \tau| \Delta^n | \tau'\rangle &= {1\over (-\del^2 + r(\tau)^2)^n}
\delta (\tau -\tau')\nn\\
&= {1\over (n-1)!} \int_0^\infty \!\!\!\!d\sigma \ \sigma^{n-1}
e^{-\sigma(-\del^2+r(\tau)^2)} \delta(\tau-\tau') \period
\end{align}
Since the exponent of the exponential in the integrand contains 
the operator $\del^2$ and the function $r^2(\tau)$ which are non-commuting, 
one needs to make use of the Baker-Campbell-Hausdorff formula up to the 
 appropriate order in the derivative expansion, so that $e^{\sigma
 \del^2}$ acts directly on $\delta(\tau-\tau')$. For more details, see  
\cite{okawa9903, hm9904}. An example of the result of such a calculation
 is
\begin{equation}
 \langle \tau | \Delta^4 | \tau \rangle = 
\frac{5}{32\,r^7} + \frac{1155\,{({r \cdot \dot{r}})}^2}
   {256\,r^{13}} - \frac{105\,({r \cdot \ddot{r}})}{128\,r^{11}} - 
  \frac{105\,\dot{r}^2}{128\,r^{11}} +\calO(\del^4) \period
\end{equation}
\subsection{Simplification procedure and the result}
As already mentioned, the procedures described above were executed by
 developing an elaborate codes for Mathematica. The number of terms 
 at the input stage is about 1000. As one performs the normal-ordering 
 and the $\ga$-matrix algebra, this number remains to be roughly of the same
 order. We then tried to simplify the results by bringing them to
appropriate standard forms via integration by parts and identification
 of the same structures with different repeated indices. This manipulation
brought the number down to about 70. 
\parsmallskip
The final step is to simplify them as much as possible via the use of 
 $SO(9)$ Fierz identities. According to the terminology of \cite{TR}, a Fierz
 identity in which $n$ of the indices of the $\ga$-matrices involved are
 {\it not} contracted among themselves is called an \lq\lq $n$-free-index"
 identity. It turns out that, including the situations that occur in 
 the next section where we examine the SUSY invariance of the effective action,
we need up to 5-free-index identities in this parlance. 
Although some class of Fierz identities are
 known in the literature and a general procedure to generate all possible
 identities was described in \cite{TR}, this was not enough for our 
purposes. The reason is two-fold: 
First, since we deal with $\theta(\tau)$ with 
 arbitrary $\tau$ dependence, $\theta, \dot{\theta},\ddot{\theta}$, etc. must
 be treated as different spinors and hence we need general forms of 
 the identities. The ones available in the literature do not cover such cases
 in sufficient generality. 
 Second, the algorithm of \cite{TR} turned out to be prohibitively 
 time-consuming when the number of free indices exceeds 3. To overcome
 this difficulty, we developed a new more efficient algorithm, which is
 described  in the Appendix A.
 None the less, one must cope with the fact that 
the number of independent Fierz identities grows like 
$\sim 5\times 2^n$ and besides the length of 
 each identity increases rapidly with $n$ as well. Consequently, 
it was a difficult task to find the right identities to be applied for 
simplification. 
\parsmallskip
However, when the right identities were applied, the result became 
 remarkably simple. The following, we believe, are the simplest 
 forms for the desired fermionic part of the 1-loop effective action
 at order 4:

\begin{align}
\Gatil_{\theta^2}^{(1)} & =
\int\!\! d\tau \ \bigg(  
  \frac{25\,v^2\,({\dot{\theta}}\theta )}{16\,r^7}
   -\frac{35\,{(r \cdot v)}^2\,({\dot{\theta}}\theta )}
     {8\,r^9}+\frac{5\,(r \cdot a)\,({\dot{\theta}}\theta )}
     {4\,r^7} \brk \hspace{4cm} +\frac{({\ddot{\theta}}{\dot{\theta}})}
     {4\,r^5}-\frac{105\,v^2\,{r_i}\,{v_j}\,
       (\theta {{\gamma }^{ij}}\theta )}{32\,r^9}+
    \frac{5\,{v_i}\,{a_j}\,(\theta {{\gamma }^{ij}}\theta )}
     {16\,r^7} 
\bigg) \comma \label{effact1} \\[1.4ex]
%
\Gatil_{\theta^4}^{(1)} & =
\int\!\! d\tau  \ \bigg(
 -\frac{25\,({\dot{\theta}}\theta )({\dot{\theta}}
        \theta )}{32\,r^7}+
    \frac{245\,{r_i}\,{v_j}\,({\dot{\theta}}\theta )\,
       (\theta {{\gamma }^{ij}}\theta )}{64\,r^9}+
    \frac{105\,{v_i}\,{v_j}\,
       (\theta {{\gamma }^{ik}}\theta )\,
       (\theta {{\gamma }^{jk}}\theta )}{128\,r^9} \brk \hspace{2cm}
   -\frac{35\,{r_i}\,{a_j}\,
       (\theta {{\gamma }^{ik}}\theta )\,
       (\theta {{\gamma }^{jk}}\theta )}{128\,r^9}-
    \frac{945\,{r_i}\,{r_j}\,{v_k}\,{v_l}\,
       (\theta {{\gamma }^{ik}}\theta )\,
       (\theta {{\gamma }^{jl}}\theta )}{128\,r^{11}}+
    \frac{5\,({\dot{\theta}}{{\gamma }^i}\theta )(
        {\dot{\theta}}{{\gamma }^i}\theta )}{32\,r^7} \brk \hspace{2cm}
   +\frac{35\,{r_i}\,{v_j}\,
       (\theta {{\gamma }^{jk}}\theta )\,
       ({\dot{\theta}}{{\gamma }^{ik}}\theta )}{64\,r^9}-
    \frac{35\,{r_i}\,{v_j}\,
       (\theta {{\gamma }^{ik}}\theta )\,
       ({\dot{\theta}}{{\gamma }^{jk}}\theta )}{32\,r^9} \bigg) \comma \\[1.4ex]
\Gatil_{\theta^6}^{(1)} & = 
\int\!\! d\tau \ \bigg(
 \frac{21\,{r_i}\,{r_j}\,({\dot{\theta}}\theta )\,
       (\theta {{\gamma }^{ik}}\theta )\,
       (\theta {{\gamma }^{jk}}\theta )}{16\,r^{11}}-
    \frac{105\,{r_i}\,{v_j}\,
       (\theta {{\gamma }^{il}}\theta )\,
       (\theta {{\gamma }^{jk}}\theta )\,
       (\theta {{\gamma }^{kl}}\theta )}{128\,r^{11}} \brk \hspace{2cm}
   -\frac{1155\,{r_i}\,{r_j}\,{r_k}\,{v_l}\,
       (\theta {{\gamma }^{im}}\theta )\,
       (\theta {{\gamma }^{jm}}\theta )\,
       (\theta {{\gamma }^{kl}}\theta )}{256\,r^{13}}-
    \frac{21\,{r_i}\,{r_j}\,
       (\theta {{\gamma }^{jl}}\theta )\,
       (\theta {{\gamma }^{kl}}\theta )\,
       ({\dot{\theta}}{{\gamma }^{ik}}\theta )}{64\,r^{11}} \bigg) \comma \\[1.4ex]
\Gatil_{\theta^8}^{(1)} &= 
 \int\!\! d\tau \ \bigg(
 -\frac{15\,(\theta {{\gamma }^{ij}}\theta )\,
       (\theta {{\gamma }^{ik}}\theta )\,
       (\theta {{\gamma }^{jl}}\theta )\,
       (\theta {{\gamma }^{kl}}\theta )}{1024\,r^{11}}-
    \frac{165\,{r_i}\,{r_j}\,
       (\theta {{\gamma }^{ik}}\theta )\,
       (\theta {{\gamma }^{jm}}\theta )\,
       (\theta {{\gamma }^{kl}}\theta )\,
       (\theta {{\gamma }^{lm}}\theta )}{512\,r^{13}} \brk \hspace{2cm}
   -\frac{2145\,{r_i}\,{r_j}\,{r_k}\,{r_l}\,
       (\theta {{\gamma }^{im}}\theta )\,
       (\theta {{\gamma }^{jm}}\theta )\,
       (\theta {{\gamma }^{kn}}\theta )\,
       (\theta {{\gamma }^{ln}}\theta )}{2048\,r^{15}} \bigg) \period
\label{effact4}
\end{align}
Here and for the rest of the article, we omit the overall common 
 factor of $N$ for simplicity. 
This constitutes one of the main results of this work. We remark that 
$\Gatil^{(1)}_{\theta^2}$ is slightly different from the one quoted 
 in a previous attempt\cite{okawa9907}, while $\Gatil^{(1)}_{\theta^8}$,
 which does not contain any derivatives, 
agrees completely with the calculation of \cite{Barrioetal}. 
Furthermore, if we drop the terms containing the derivatives of $\theta$ and 
 the acceleration $a$, our result coincides (up to an overall constant)
 with the previous calculations \cite{kraus, mcarthur, Hyunetal} 
for this very special configuration.  
As for comparison with 11-dimensional supergravity, we shall make a remark
in the final section. 
\parsmallskip
In view of the fact that the calculation consisted of enormous number of steps
 and that we deal with fully off-shell configurations, 
the agreement cited above for special cases should only be regarded as
 a strong but not decisive evidence for the correctness of our calculation:
It is certainly desirable to have an independent check. 
In the next section, we shall perform such a test by computing the 
effective SUSY transformation laws for the effective action and demonstrate
 that indeed the effective action shown above is invariant under these
 transformations. 
\section{Effective SUSY Transformation Laws and Invariance of the 
Effective Action}
\subsection{A brief review of the SUSY Ward identity}
Since the calculation of the effective SUSY transformation laws will be
 based on the supersymmetric Ward identity in the background 
 gauge derived in a previous 
 publication \cite{Kaz-Mura}, let us briefly review its salient features.
\parsmallskip
The derivation of the SUSY Ward identity in the path-integral formalism
 is more or less a textbook matter, except for two non-trivialities. 
One stems from the intertwining of the SUSY and BRST symmetries. 
In order to bring the SUSY Ward identity in a useful form, one must make 
judicious uses of the BRST Ward identities. The other complication is 
 due to the two different origins of the $B_m$ dependence, one from the 
separation of the background and the quantum fluctuation, namely
 $X_m = (1/g)B_m + Y_m$, and the other from the gauge-fixing function
 $\Gtil = -\del_\tau \Atil + i\com{B^m}{X_m} $. In order to obtain the 
 correct Ward identity, one must carefully distinguish these two origins.
For more details, see \cite{Kaz-Mura}. 
\parsmallskip
Through the procedure outlined above, one obtains the SUSY Ward
 identity in the form where the effective SUSY transformation laws
 can be read off in closed forms.  Adapted to the source-probe situation 
 under consideration, it takes the form 
\begin{equation}
0 = \int\!\! d\tau 
\left( \Delta_\ep r_m (\tau)
{\delta \Gatil \over \delta r_m(\tau)}
 +  \Delta_\ep \theta_{\al}(\tau)
{\delta\Gatil \over \delta \theta_{\al}(\tau) } \right)\comma 
\end{equation}
where the effective SUSY transformation laws are given by 
\begin{align}
\Delta_\ep r_m (\tau) &= \int\!\! d\tau' \  T ^{-1}_{m,n}(\tau',\tau)
(\langle \delta_\ep \yhat_n(\tau') \rangle +
 \langle \delta_{\rm B} \yhat_n(\tau') \calO_\ep\rangle )\comma \label{delr} \\
\Delta_\ep \theta_{\al}(\tau) &=
 \langle \delta_\ep \psihat_{\al}(\tau)\rangle 
 -\langle \delta_{\rm B} \psihat_{\al}(\tau)\calO_\ep\rangle \nn\\
& - \int\!\! d\tau' d\tau'' \  T ^{-1}_{m,n}(\tau'',\tau') 
\langle \delta_{\rm B} \psihat_{\al}(\tau) \calO_{n}(\tau') \rangle \nn\\
& \qquad  \times
 \langle \delta_\ep \yhat_{m}(\tau'') + \delta_{\rm B} \yhat_{m}(\tau'')\calO_\ep
\rangle \period \label{delth}
\end{align}
In the expressions above, $\delta_\ep$ is the SUSY variation (\ref{treesusy}),
$\delta_{\rm B}$ is the BRST variation (\ref{brstr}), 
$\yhat_m$ and $\psihat_\al$ respectively denote
$gY_{11}$ and $g\Psi_{\al,11}$ (\ie the diagonal fluctuation of $r_m$ and 
 $\theta_\al$ respectively),
 $\langle \quad \rangle$ expresses the the expectation 
 value, and the operators $\calO_\ep$, $\calO_m$ and the kernel
 $T_{n,m}(\tau,\tau')$ are given as shown below:
\begin{align}
\calO_\ep &= -i\trace \!\int\!\! d\tau  \ \Cbar \delta_\ep (-\del_\tau \Atil 
 + i\com{B^m}{X_m} ) \comma \\
\calO_{m} &= \sum_{I=2}^{N+1}
\left( \Cbar_{I1}Y_{m,1I}-\Cbar_{1I}Y_{m,I1}\right) \comma \\
 T _{n,m}(\tau,\tau') &\equiv  
\delta_{mn}\delta (\tau-\tau') 
 -\langle \delta_{\rm B} \yhat_{n}(\tau') \calO_{m}(\tau)\rangle \period
\end{align}
Note that these quantum-corrected transformation laws 
are much more involved than those at the tree level. Even the 
 linear law $\delta_\ep X_m =-i\ep^T\ga_m\Theta$ for the bosonic
 field gets modified non-trivially contrary to naive expectation. 
It is not simply the expectation value of the linear law itself because
 what is relevant is the effective law that acts on $\Gatil$. 
\subsection{Calculation of the effective SUSY transformation laws}
Actual evaluation of $\Delta r_m$ and $\Delta \theta_\al$ at 1-loop to
 the relevant order is essentially straight-forward but extremely cumbersome.
Since similar calculations performed at order 2 were fully displayed 
 in the previous work\cite{Kaz-Mura}, we shall only give a sketch of 
the general  procedure for the present case.
\parsmallskip
When the variations $\delta_\ep$ and $\delta_{\rm B}$ are performed and the 
 definitions of the composite operators $\calO_\ep$ and $\calO_m$ are
 substituted into (\ref{delr}) and (\ref{delth}), it is not difficult to 
 see that, at 1-loop, terms contributing to
 $\Delta r_m$ and $\Delta \theta_\al$ 
 can be expressed as Feynman diagrams composed of products of  
 tree-level 2-point functions for various fields. These are of course
 computed from the quadratic part of the gauge-fixed action already
 displayed in (\ref{quadaction}). The only complication is that 
 we must disentangle the mixings among fields, which, due to the presence
 of the fermionic background, includes those between bosonic and fermionic 
 fields. The results for the non-trivial 2-point functions are given 
 in the Appendix B. 
\parsmallskip
Although most of these diagrams produce local expressions as desired, 
 but there exist some which give non-local contributions. 
For example, a term contributing to $\Delta r_m$ is of the form
\begin{align}
 g^2\!\int\!\!d\tau^\prime d\tau^{\prime\prime} \ 
\lk C(\tau)\Cbar^\ast(\taupp)\rk 
\lk C_{11}(\taupp)\dot{\Cbar}_{11}(\taup)\rk
\lk \left( -\dot{\tilA \ }\! + ir_n Y_n \right)
(\taupp)Y^\ast_m(\tau)\rk  
\epsilon_\beta  \theta_\beta(\taup) \comma
\end{align}
where the second factor  $\lk C_{11}(\taupp)\dot{\Cbar}_{11}(\taup)\rk$
going like $1/\del$ produces unwanted non-locality. Fortunately, 
 when one isolates all the terms of this sort, one notices that they 
 precisely cancel in the SUSY Ward identity due to the
 following BRST Ward identity:
\begin{align}
\lk \delta_{\rm B} \calQ(\taupp) \rk &= 
-\int\!\! d\tau 
\left(
\frac{\delta \widetilde{\Gamma}}{\delta r_m(\tau)} 
\lk \delta_{\rm B}\widehat{y}_m(\tau)  \calQ(\taupp)  \rk  
+
\frac{\delta \widetilde{\Gamma}}{\delta \theta_\alpha(\tau)}
\lk \delta_{\rm B}\widehat{\psi}_\alpha(\tau) \calQ(\taupp)  \rk  
\right) \ , 
\label{brst}\\
\calQ(\taupp)  & \equiv -i \Cbar(\taupp) 
\left( - \dot{\tilA^\ast} - ir_n Y^\ast_n \right)\!(\taupp)
+
i \Cbar^\ast (\taupp)
 \left( -\dot{\tilA \ }\! + ir_n Y_n \right)\!(\taupp) \ . 
\end{align}
We omit the details of the demonstration. 
\parsmallskip
Just as for the effective action, the calculation of the relevant diagrams
for the SUSY transformation laws was performed with the aid of Mathematica.
Even after various simplification procedures the results are quite 
 complicated, with the longest expression, $\Delta \theta_\al$ 
 with 4 $\theta$'s, consisting of 56 terms. So we relegate them to the 
Appendix C  in order not to interrupt the flow of the main text. 
A glance at this result would convince one that, when we prove 
the invariance of effective action under these transformations in the 
sequel, calculations of both the effective action and the SUSY 
transformations must be correct. 
\subsection{SUSY invariance of the effective action}
Now we come to the last stage of this work, the check of the invariance
 of the effective action under the SUSY transformation laws just computed.
 Logically what we wish to demonstrate
 is quite simple: The tree level SUSY variation of the 1-loop effective action
 and the 1-loop level SUSY variation of the tree action must cancel. 
The difficulty is again that above $\calO(\theta^4)$
 we need judicious applications of various Fierz identities to effect 
such cancellations. The required Fierz identities are more involved than 
 those used in the process of simplifying the effective action itself, 
 since we have one more independent spinor, namely the SUSY transformation
 parameter $\ep_\al$. Below we give a sketch of how such cancellations
 take place, choosing the case involving $\Gatil^{(1)}_{\theta^4}$
 as an example. 
\parsmallskip
For the structure with one $\ep$ and 3 $\theta$'s, what we wish to prove is
\begin{eqnarray}
(\delta \Gatil)_{\ep,\theta^3} =\delta^{(0)}_r \Gatil^{(1)}_{\theta^2} + 
+\delta^{(0)}_\theta \Gatil^{(1)}_{\theta^4}+\delta^{(1)}_r \Gatil^{(0)}
+\delta^{(1)}_\theta \Gatil^{(0)} =0 \period
\end{eqnarray}
Substituting the 
SUSY transformation laws and bring the result into certain standard
 forms by integration by parts, $\delta \Gatil$ turned out to 
 contain 40 terms. They are classified into two groups, one with 3 free
 indices and the rest with 1 free index.  We then apply three independent 
 3-free-index Fierz identities to reduce the terms in the first group 
 down to 1-free-index type. Since the identities used are somewhat complicated,
 let us display only one of them as an example:
\begin{align}
(\epsilon {{\gamma }^{{a_1}k}}
    {\dot{\theta}})\,
   (\theta {{\gamma }^{{a_1}ij}}
    \theta ) = 
   \and -(\epsilon 
       {{\gamma }^{{a_1}jk}}
       {\dot{\theta}})\,
      (\theta {{\gamma }^{{a_1}i}}
       \theta )
    +(\epsilon {{\gamma }^{{a_1}ik}}
       {\dot{\theta}})\,
      (\theta {{\gamma }^{{a_1}j}}
       \theta )
    +(\epsilon {{\gamma }^k}
       {\dot{\theta}})\,
      (\theta {{\gamma }^{ij}}
       \theta ) \brkeq 
    -(\epsilon {{\gamma }^j}
       {\dot{\theta}})\,
      (\theta {{\gamma }^{ik}}
       \theta )
    +(\epsilon {{\gamma }^i}
       {\dot{\theta}})\,
      (\theta {{\gamma }^{jk}}
       \theta )
    -(\epsilon {\dot{\theta}})\,
      (\theta {{\gamma }^{ijk}}
       \theta ) \brkeq 
    +4(\epsilon {{\gamma }^{ij}}
       \theta )\,
      ({\dot{\theta}}{{\gamma }^k}
       \theta )
    +2(\epsilon {{\gamma }^k}\theta 
       )\,({\dot{\theta}}
       {{\gamma }^{ij}}\theta )
    +2(\epsilon {{\gamma }^j}\theta 
       )\,({\dot{\theta}}
       {{\gamma }^{ik}}\theta )
     \brkeq -2(\epsilon 
       {{\gamma }^i}\theta )\,
      ({\dot{\theta}}
       {{\gamma }^{jk}}\theta )
    +2(\epsilon \theta )\,
      ({\dot{\theta}}
       {{\gamma }^{ijk}}\theta )
    -2(\epsilon {{\gamma }^{{a_1}}}
       \theta )\,
      ({\dot{\theta}}
       {{\gamma }^{{a_1}ijk}}\theta 
       ) \brkeq 
    -2({\dot{\theta}}\theta )\,
      (\epsilon {{\gamma }^j}\theta 
       )\,{{\delta }_{ik}}
    +(\epsilon {{\gamma }^{{a_1}}}
       {\dot{\theta}})\,
      (\theta {{\gamma }^{{a_1}j}}
       \theta )\,{{\delta }_{ik}}
    +2(\epsilon \theta )\,
      ({\dot{\theta}}{{\gamma }^j}
       \theta )\,{{\delta }_{ik}}
     \brkeq -2(\epsilon 
       {{\gamma }^{{a_1}}}\theta )\,
      ({\dot{\theta}}
       {{\gamma }^{{a_1}j}}\theta )\,{{\delta }_{ik}}
    +2({\dot{\theta}}\theta )\,
      (\epsilon {{\gamma }^i}\theta 
       )\,{{\delta }_{jk}}
    -(\epsilon {{\gamma }^{{a_1}}}
       {\dot{\theta}})\,
      (\theta {{\gamma }^{{a_1}i}}
       \theta )\,{{\delta }_{jk}}
     \brkeq -2(\epsilon \theta )\,
      ({\dot{\theta}}{{\gamma }^i}
       \theta )\,{{\delta }_{jk}}
    +2(\epsilon {{\gamma }^{{a_1}}}
       \theta )\,
      ({\dot{\theta}}
       {{\gamma }^{{a_1}i}}\theta )\,{{\delta }_{jk}} \period
\end{align}
After this reduction, $\delta\Gatil$ consists of 29 terms, all
 with one free index only. To this we apply the following 
relatively simple 1-free-index Fierz identities:
\begin{align}
&({\rm i})\quad (\epsilon {{\gamma }^{{a_1}}}
    {\dot{\theta}})\,
   (\theta {{\gamma }^{{a_1}i}}
    \theta ) =
   2({\dot{\theta}}\theta )\,
      (\epsilon {{\gamma }^i}\theta 
       )-(\epsilon 
       {{\gamma }^{{a_1}i}}\theta )\,({\dot{\theta}}
       {{\gamma }^{{a_1}}}\theta ) \nn\\
& \hspace{5cm}  -2(\epsilon \theta )\,
      ({\dot{\theta}}{{\gamma }^i}
       \theta )
    +(\epsilon {{\gamma }^{{a_1}}}
       \theta )\,
      ({\dot{\theta}}
       {{\gamma }^{{a_1}i}}\theta )\comma  \nn\\
&({\rm ii}) \quad (\epsilon {{\gamma }^{{a_1}}}
    {\ddot{\theta}})\,
   (\theta {{\gamma }^{{a_1}i}}
    \theta ) =
   2({\ddot{\theta}}\theta )\,
      (\epsilon {{\gamma }^i}\theta 
       )-(\epsilon 
       {{\gamma }^{{a_1}i}}\theta )\,({\ddot{\theta}}
       {{\gamma }^{{a_1}}}\theta )\nn\\
& \hspace{5cm} -2(\epsilon \theta )\,
      ({\ddot{\theta}}{{\gamma }^i}
       \theta )
    +(\epsilon {{\gamma }^{{a_1}}}
       \theta )\,
      ({\ddot{\theta}}
       {{\gamma }^{{a_1}i}}\theta ), \nn\\
&({\rm iii}) \quad (\epsilon {{\gamma }^{{a_1}}}\theta 
    )\,({\dot{\theta}}
    {{\gamma }^{{a_1}i}}
    {\dot{\theta}}) = 
   -2({\dot{\theta}}\theta )\,
      (\epsilon {{\gamma }^i}
       {\dot{\theta}})
    +(\epsilon {{\gamma }^{{a_1}i}}
       {\dot{\theta}})\,
      ({\dot{\theta}}
       {{\gamma }^{{a_1}}}\theta ) \nn\\
& \hspace{5cm} +2(\epsilon 
       {\dot{\theta}})\,
      ({\dot{\theta}}{{\gamma }^i}
       \theta )
    +(\epsilon {{\gamma }^{{a_1}}}
       {\dot{\theta}})\,
      ({\dot{\theta}}
       {{\gamma }^{{a_1}i}}\theta ), \nn\\
&({\rm iv}) \quad (\epsilon {{\gamma }^{{a_1}{a_2}}}
    {\dot{\theta}})\,
   ({\dot{\theta}}
    {{\gamma }^{{a_1}{a_2}i}}\theta 
    ) =   -2
     ({\dot{\theta}}\theta )\,
      (\epsilon {{\gamma }^i}
       {\dot{\theta}})
    -2(\epsilon {{\gamma }^{{a_1}i}}
       {\dot{\theta}})\,
      ({\dot{\theta}}
       {{\gamma }^{{a_1}}}\theta ) \nn\\
& \hspace{5cm} -6(\epsilon 
       {\dot{\theta}})\,
      ({\dot{\theta}}{{\gamma }^i}
       \theta )
    -2(\epsilon {{\gamma }^{{a_1}}}
       {\dot{\theta}})\,
      ({\dot{\theta}}
       {{\gamma }^{{a_1}i}}\theta )\comma  \nn\\
&({\rm v})\quad (\epsilon {{\gamma }^{{a_1}{a_2}}}
    {\dot{\theta}})\,
   ({\dot{\theta}}
    {{\gamma }^{{a_1}{a_2}i}}
    {\dot{\theta}}) = 
  -2(\epsilon {{\gamma }^{{a_1}}}
     {\dot{\theta}})\,
    ({\dot{\theta}}
     {{\gamma }^{{a_1}i}}
     {\dot{\theta}}) \period \nn
\end{align}
Then $(\delta \Gatil)_{\ep,\theta^3} $
 can be brought to the following form, in which 
 the 4 fermions involved are all different:
\begin{align}
 (\delta \Gatil)_{\ep,\theta^3} = 
\int\!\! d\tau \  \bigg(
 \and \ {\frac{15\,i\,{r_{{i_1}}}\,
        ({\ddot{\theta}}
         {\dot{\theta}})\,
        (\epsilon 
         {{\gamma }^{{i_1}}}\theta )
        }{16\,{r^7}}}
   +{\frac{15\,i\,{r_{{i_1}}}\,
        ({\ddot{\theta}}\theta )\,
        (\epsilon 
         {{\gamma }^{{i_1}}}
         {\dot{\theta}})}{16\,{r^7}}
      }-{\frac{5\,i\,{r_{{i_1}}}\,
        (\epsilon 
         {{\gamma }^{{a_1}}}
         {\ddot{\theta}})\,
        ({\dot{\theta}}
         {{\gamma }^{{a_1}{i_1}}}
         \theta )}{16\,{r^7}}}
    \brkeq +{\frac{5\,i\,
        {r_{{i_1}}}\,
        (\epsilon 
         {{\gamma }^{{a_1}{a_2}}}
         {\ddot{\theta}})\,
        ({\dot{\theta}}
         {{\gamma }^
           {{a_1}{a_2}{i_1}}}\theta 
         )}{32\,{r^7}}}
   +{\frac{5\,i\,{r_{{i_1}}}\,
        (\epsilon {\dot{\theta}})\,
        ({\ddot{\theta}}
         {{\gamma }^{{i_1}}}\theta )
        }{16\,{r^7}}}
   +{\frac{5\,i\,{r_{{i_1}}}\,
        (\epsilon \theta )\,
        ({\ddot{\theta}}
         {{\gamma }^{{i_1}}}
         {\dot{\theta}})}{16\,{r^7}}
      } \brkeq 
   +{\frac{5\,i\,{r_{{i_1}}}\,
        (\epsilon 
         {{\gamma }^{{a_1}}}
         {\dot{\theta}})\,
        ({\ddot{\theta}}
         {{\gamma }^{{a_1}{i_1}}}
         \theta )}{16\,{r^7}}}
   +{\frac{5\,i\,{r_{{i_1}}}\,
        (\epsilon 
         {{\gamma }^{{a_1}}}\theta )
         \,({\ddot{\theta}}
         {{\gamma }^{{a_1}{i_1}}}
         {\dot{\theta}})}{16\,{r^7}}
      } \bigg)
\period \label{susyinv}
\end{align}
Finally, by a Fierz identity of the type described in the Appendix A, this 
 vanishes identically. 
\parsmallskip
The vanishing of $(\delta \Gatil)_{\ep,\theta} $ can be proved similarly
 with much less effort. On the other hand, for $(\delta \Gatil)_{\ep,\theta^5}$
 and $(\delta \Gatil)_{\ep,\theta^7}$ we encounter a considerable 
difficulty: We needed up to 5-free-index identities which are in general
 quite complicated\footnote{The longest 5-free-index identity
we used consists of 109 terms.} and moreover for expressions with
 more than six spinors there are many possibilities to choose the 
 four among them to which to apply the Fierz. If the choice of which 
 identity to use or which spinors to act on is not
 appropriate, the result immediately 
becomes more complicated, by dozens of terms,
 than the one prior to the application. For these reasons, 
even after considerable amount of trial end error efforts  we could 
 not fully identify all the Fierz identities responsible for the vanishing of 
 the variation. To overcome this difficulty, we took the following
 strategy: First simplify the expression as much as possible by the 
 judicious application of various Fierz identities. Then, when it gets reduced 
 to a sufficiently simple form,  check if it vanishes  \lq\lq numerically" by
 the use of explicit representations of the $SO(9)$ $\ga$-matrices. 
In this way, we succeeded in checking the full SUSY invariance of 
 our effective action. 
\section{Summary and Discussions}
In this paper, as a necessary step toward clarification of 
 the power of supersymmetry in Matrix theory, we have performed 
 a complete calculation of the 1-loop 
effective action for {\it arbitrary} off-shell
 trajectory and spin degrees of freedom for a probe D-particle 
at order 4 in the derivative expansion. Although the calculation was 
 quite involved in the intermediate stages, the result presented in 
(\ref{effact1}) $\sim$ (\ref{effact4}) turned out to be  remarkably simple. 
Further, for an independent check of this result as well as for its own 
interest, we have computed the quantum-corrected SUSY transformation 
 laws to the appropriate order  based on the SUSY Ward identity developed 
previously\cite{Kaz-Mura}. This computation was again extremely cumbersome 
 and we obtained  rather complicated results shown in Appendix C. 
To test the invariance of the effective action under these transformations,
 it was crucial to apply a series of judicious Fierz identities, many of 
 which had not been known. We developed a new efficient algorithm and 
 generated the necessary identities. With the aid of these identities and 
 some numerical computations, we succeeded in checking the desired 
invariance. 

An obvious important question about our result for the effective action
is whether it agrees with the supergravity calculation. Unfortunately, 
 at present time this question cannot be answered for the following reason. 
 A relevant supergravity calculation was performed in \cite{HyunetalPRD}
 up to $\calO(\theta^2)$, adapting a technique developed for the 
 case of 11-dimensional supermembrane in \cite{deWitetal} to the 
 same order. Subsequently it was 
 argued in \cite{okawa9907} that by appropriate re-definitions 
 of the fields $r_m(\tau)$ and $\theta_\al(\tau)$ such an effective action 
 can be made to agree with the Matrix theory calculation to that 
order\footnote{
These re-definitions were not unique since any difference at 
$\calO(\theta^4)$ was neglected.}. In order to check our complete result
 including all the spin effects at order 4, one needs a calculation 
 on the supergravity side up to $\calO(\theta^8)$. Judging from the 
existing 
calculation at $\calO(\theta^2)$ level\cite{deWitetal},
 which contains some ambiguities left unresolved, this appears 
 to be quite a difficult task. Thus, assuming that the agreement 
 with the supergravity calculation would persist, our result stands 
 as a prediction (up to field re-definitions) until such a challenge 
 will have been met. 
\parsmallskip
We conclude by recalling  the prime motivation that prompted 
 this work, namely the study of the power of supersymmetry in 
Matrix theory.
 What is  most curious is to see how much of the highly non-trivial yet
 remarkably simple structure of the completely off-shell 
order 4 effective action we computed is determined by supersymmetry alone. 
This requires enumerating the most general form of the effective 
 action at this order and study how the coefficients of independent 
 structures are restricted by the requirement of SUSY invariance. 
Such a work is now underway\cite{Kaz-Mura3} and we hope to communicate
 the result elsewhere. 
\par\bigskip\noindent
{\large\bf Acknowledgment}\par\smallskip\noindent
The research of Y.K. is supported in part by 
Grant-in-Aid for Scientific Research
on Priority Area \#707 \lq\lq Supersymmetry and Unified Theory of 
 Elementary Particles" No.~10209204 and 
 Grant-in-Aid for Scientific Research (B) 
No.~12440060, while that of T.M is supported in part by 
the Japan Society for Promotion of Science under the Predoctoral
Research Program No.~12-9617, 
all from the Japan Ministry of Education, Culture, Sports, Science
 and Technology. 
\newpage
\setcounter{equation}{0}
\renewcommand{\theequation}{A.\arabic{equation}}
\noindent
{\Large\bf Appendix A:} {\large\bf \quad
A new efficient algorithm for 
 generating $SO(9)$ Fierz identities}
\parbigskipn
In this appendix, we describe a new efficient algorithm for generating 
 the $SO(9)$ Fierz identities and make  a  remark on
the existence of a class of identities which are often overlooked. 
\parsmallskip
Let us define  two types of four fermion structures 
 $f_n^m$ and $g_n^m$, with $n$ indices 
 contracted and $m$ indices free,  as follows\footnote{Labels which 
 distinguish the various ways the free indices are distributed 
 are suppressed.} :
\begin{align}
f_n^m &= (\lam_1^T \ga^{a_1 \ldots a_n i_1 \ldots i_k}
\lam_2)(\lam_3^T \ga^{a_n \ldots a_1 i_{k+1} \ldots i_m} \lam_4) \comma \\
g_n^m &= (\lam_2^T \ga^{a_1 \ldots a_n i_1 \ldots i_k}
\lam_3)(\lam_4^T \ga^{a_n \ldots a_1 i_{k+1} \ldots i_m} \lam_1) \period
\end{align}
The generic Fierz identities relate these two types of fermion bilinears. 
(In general, some of the free indices may be carried by factors of 
 Kronecker $\delta$'s such as $\delta^{i_1i_2}$,etc.)
In \cite{TR} a systematic procedure to generate such identities was described.
A part of this algorithm requires generation of tensor product 
 identities for $\ga$-matrices and this turned out to be progressively 
 time-consuming as the number of free indices increases. Besides,
 the set of Fierz identities so obtained are highly redundant, including 
 the repetition of relations already obtained for lower number of 
free indices. 
\parsmallskip
The basic idea of our new algorithm is to make full use of the 
 Fierz identities for $m$-free-index in the calculation 
of the $(m+1)$-free-index case, \ie it is an inductive algorithm. 
Suppose we already have all the identities with up to  $m$ free indices. 
Concentrating on the $\ga$-matrix structure, we have $m+1$ $f$-type 
 structures of the form 
\begin{equation}
\ga^{\ind{a}{n} \ind{i}{k}} _{ab}\ga^{\indr{a}{n} i_{k+1}\ldots i_m}_{cd}
\end{equation}
expressed in terms of the $g$-type structures.
  Now we try to add another free index $j$ to this relation. Clearly
 there are 4 places to insert $\ga^j$. So we generate $4(m+1)$ relations. 
This is considerably smaller than $4^{m+1}$ relations generated by the 
 method  of \cite{TR}. 
\parsmallskip
For example, adding $\ga^j$ to the left-most position, we get, in 
 an obvious tensor product notation, 
\begin{align}
& \ga^j \ga^{\ind{a}{n} \ind{i}{k}}\otimes 
\ga^{\indr{a}{n} i_{k+1}\ldots i_m} \nn\\
& \qquad = (-1)^n \ga^{a_1 \ldots a_n ji_1 \ldots i_k} \otimes
\ga^{\indr{a}{n} i_{k+1}\ldots i_m} \nn\\
 & \qquad\quad  + n\ga^{\ind{a}{{n-1}} \ind{i}{k}}
\otimes \ga^{\indr{a}{{n-1}} j i_{k+1}\ldots i_m} \nn\\
& \qquad \qquad+ \sum_l (-1)^{n+l-1}\delta^{ji_l} \ga^{\ind{a}{n} i_1 \ldots
 \hat{i}_l\ldots i_k} \otimes \ga^{\indr{a}{n} i_{k+1}\ldots i_m}
\comma 
\end{align}
where $\hat{i}_l$ means that it is deleted. The first term contains 
 $m+1$ free indices, while the second sum contains structures with 
 $m-1$ indices, which can be reduced into $g$-type structures by 
 the relations already computed. 
\parsmallskip
Now on the right hand side of the original $m$-free-index Fierz relations, 
we have $g$-type structures of the form ($**$ stands for a set of 
indices) 
\begin{equation}
\sum C \ga^{**} _{bc} \ga^{**}_{da}\period
\end{equation}
When $\ga^j$ is added in the manner above, this becomes 
\begin{equation}
\sum C \ga^{**} _{bc} (\ga^{**}\ga^j)_{da}\comma 
\end{equation}
which can easily be computed. This produces $g$-type structures with 
 one more or one less free indices on the $\ga$'s. Equating the left and 
 the right hand sides, we produce a $(m+1)$-free-index identity. 
\parsmallskip
Now we make a cautionary remark in enumerating all possible 
 Fierz identities when all the spinors involved are distinct. 
The Fierz identities discussed so far (and in \cite{TR}) 
 are the ones where the order of the spinors are cyclically rotated. 
Schematically, the relation is of the type 
$(\lam_1\ga \lam_2)(\lam_3\ga\lam_4) \rightarrow 
(\lam_2\ga \lam_3)(\lam_4\ga\lam_1)$. Further cyclic rotation 
 of course does not produce new identities. However, there is an additional
 structure $(\lam_1\ga \lam_3)(\lam_2\ga\lam_4)$, which cannot be reached by
 cyclic rotations from the original. Therefore, one must also add the 
 transformation of the type $(\lam_1\ga \lam_2)(\lam_3\ga\lam_4)
\rightarrow (\lam_1\ga \lam_3)(\lam_2\ga\lam_4)$.  One should further be
 aware that 
 in general these two classes of identities may contain redundancy. 
So the correct procedure is to generate all the identities of both classes
 and then re-solve them as coupled equations to find the truly independent
 and complete set of identities. For example, the identity which states
the vanishing of (\ref{susyinv}) was obtained only through this 
 procedure. 
\parsmallskip
In this way, starting from the 0-free-index 
 identities given in \cite{TR} we have generated all the independent 
Fierz identities up to and including
 4 free indices and a part of 5-free-index ones. Unfortunately, the 
 result is too space-filling to be displayed in this article. 

\parbigskipn
\setcounter{equation}{0}
\renewcommand{\theequation}{B.\arabic{equation}}
\noindent
{\Large\bf Appendix B:} {\large\bf \quad Two-point functions needed for 
 the calculation of SUSY transformation laws}
\parbigskipn
We list the non-trivial two-point functions which serve as 
 the basic elements in the calculation of SUSY transformation laws. 
In the expressions below, $\Xi_{\al M}$ and $\Phi_M$ are the 
$(1+9)$-component fermionic  and bosonic  vectors introduced 
 in (\ref{XiPhi}), and $D^{-1}_{B MN}$ and $D^{-1}_{F \al\be}$ are 
the bosonic and 
 fermionic propagator matrices  described in (\ref{Bprop}) and 
(\ref{Fprop}) respectively, with 
 the component indices $M,N$ etc. (running from $0$ to $9$)
 displayed explicitly for clarity. Further, 
 $D_{\rm B}^{\dagger -1}$ and $D_{\rm F}^{\dagger -1}$,
which were not explicitly given in the main text, are defined as
\begin{align}
D_{{\rm B} NM}^{\dagger -1}&= 
\left( 
\barr{cc}
\Delta^{-1}&
2iv^m\\
-2iv^n&
 \Delta^{-1} \delta_{nm}  
\earr
\right)^{-1} \nn \\
&= 
\left( 
\barr{cc}
\Delta (1 - 4v^\ell\Delta v^\ell\Delta)^{-1} &
-2i\Delta (1 - 4v^\ell\Delta v^\ell\Delta)^{-1} v^m \Delta \\
2i\Delta v^n \Delta(1 - 4v^\ell\Delta v^\ell\Delta)^{-1} &
 \Delta (\delta_{nm} -4v^n\Delta v^m\Delta )^{-1}
\earr
\right) ~,
\end{align}
\begin{align}
 D_{\rm F}^{\dagger -1} = (\p - \dirac{r}_{ij})^{-1}=
-(\p + \dirac{r}_{ij})
(1 - \Delta\dirac{v}_{ij})^{-1}\Delta ~.
\end{align}

With these notations, the basic two point functions are given by 
\begin{align}
 \lk \Phi^\ast_M(\tau) \Phi_N(\taup) \rk & = 
\bra{\tau}
(\delta_{ML}-
D_{{\rm B} MK}^{-1}   
 \Xi_{K \rho}
D_{{\rm F} \rho\sigma}^{-1}   
\Xi_{\sigma L}
)^{-1} D_{{\rm B} LN}^{-1}
\ket{\taup}   \ , \\[1.2ex]
 \lk \Psi^\ast_\alpha(\tau) \Phi_M(\taup) \rk & = 
\bra{\tau}
-(\delta_{ML}-
D_{{\rm B} MK}^{-1}   
 \Xi_{K \rho}
D_{{\rm F} \rho\sigma}^{-1}   
\Xi_{\sigma L}
)^{-1} D_{{\rm B} LP}^{-1}
\Xi_{P \lambda} D_{{\rm F} \lambda\alpha}^{-1} 
\ket{\taup}
   \ , \\[1.2ex]
 \lk \Psi^\ast_\alpha(\tau) \Psi_\beta(\taup) \rk & = 
\bra{\tau}
(\delta_{\alpha\lambda}-
D_{{\rm F} \alpha\rho}^{-1}   
 \Xi_{\rho K}
D_{{\rm B} KL}^{-1}   
\Xi_{L \lambda}
)^{-1} 
D_{{\rm F} \lambda\beta}^{-1} 
\ket{\taup}
\ , \\[1.2ex]
 \lk \Phi^\ast_N(\tau) \Psi_\beta(\taup) \rk & = 
\bra{\tau}
-(\delta_{\alpha\lambda}-
D_{{\rm F} \alpha\rho}^{-1}   
 \Xi_{\rho K}
D_{{\rm B} KL}^{-1}   
\Xi_{L \lambda}
)^{-1} 
D_{{\rm F} \lambda\sigma}^{-1}
\Xi_{\sigma P} D_{{\rm B} PN}^{-1}   
\ket{\taup}
\ ,
\end{align}
while their \lq\lq conjugates" take the form 
\begin{align}
 \lk \Phi_M(\tau) \Phi^\ast_N(\taup) \rk & = 
\bra{\tau}
(\delta_{ML}-
D_{{\rm B} MK}^{\dagger -1}   
 \Xi_{K \rho}
D_{{\rm F} \rho\sigma}^{\dagger -1}   
\Xi_{\sigma L}
)^{-1} D_{{\rm B} LN}^{\dagger  -1}
\ket{\taup}   \ , \\[1.2ex]
 \lk \Psi_\alpha(\tau) \Phi^\ast_M(\taup) \rk & = 
\bra{\tau}
(\delta_{ML}-
D_{{\rm B} MK}^{\dagger -1}   
 \Xi_{K \rho}
D_{{\rm F} \rho\sigma}^{\dagger -1}   
\Xi_{\sigma L}
)^{-1} D_{{\rm B} LP}^{\dagger -1}
\Xi_{P \lambda} D_{{\rm F} \lambda\alpha}^{\dagger -1} 
\ket{\taup}
   \ , \\[1.2ex]
 \lk \Psi_\alpha(\tau) \Psi^\ast_\beta(\taup) \rk & = 
\bra{\tau}
(\delta_{\alpha\lambda}-
D_{{\rm F} \alpha\rho}^{\dagger -1}   
 \Xi_{\rho K}
D_{{\rm B} KL}^{\dagger -1}   
\Xi_{L \lambda}
)^{-1} 
D_{{\rm F} \lambda\beta}^{\dagger -1} 
\ket{\taup}
\ , \\[1.2ex]
 \lk \Phi_N(\tau) \Psi^\ast_\beta(\taup) \rk & = 
\bra{\tau}
(\delta_{\alpha\lambda}-
D_{{\rm F} \alpha\rho}^{\dagger -1}   
 \Xi_{\rho K}
D_{{\rm B} KL}^{\dagger -1}   
\Xi_{L \lambda}
)^{-1} 
D_{{\rm F} \lambda\sigma}^{\dagger -1}
\Xi_{\sigma P} D_{{\rm B} PN}^{\dagger -1}   
\ket{\taup}
\ .
\end{align}
Of course in the actual calculation, we must expand them to the 
 appropriate order in the derivative expansion. 
\newpage
\setcounter{equation}{0}
\renewcommand{\theequation}{C.\arabic{equation}}
\noindent
{\Large\bf Appendix C:} {\large\bf \quad  Quantum-corrected SUSY
 transformation laws}
\parbigskipn
In this appendix we display the 1-loop corrections to the  supersymmetry
 transformation laws needed for checking the invariance
 of the order 4 effective action. The results are classified according to the 
  number of $\theta$'s in excess of the tree-level laws. 
\shead{Corrections at ${\cal O}(\theta^0)$:}
\begin{align}
\Delta_\epsilon r^m =
 \and -\frac{5\,i\,g^2\,{v_i}\,{v_m}\,
       (\epsilon {{\gamma }^i}\theta )}{2\,r^7}+
    \frac{5\,i\,g^2\,v^2\,(\epsilon {{\gamma }^m}\theta )}
     {4\,r^7}-\frac{35\,i\,g^2\,{(r \cdot v)}^2\,
       (\epsilon {{\gamma }^m}\theta )}{8\,r^9} \brkeq 
   +\frac{5\,i\,g^2\,(r \cdot a)\,
       (\epsilon {{\gamma }^m}\theta )}{4\,r^7}+
    \frac{5\,i\,g^2\,(r \cdot v)\,
       (\epsilon {{\gamma }^m}{\dot{\theta}})}{4\,r^7}-
    \frac{i\,g^2\,(\epsilon {{\gamma }^m}{\ddot{\theta}})}
     {4\,r^5} .
\end{align}
\begin{align}
\Delta_\epsilon\theta_\alpha =
 \and -\frac{5\,i\,g^2\,v^2\,{v_i}\,
       {{(\epsilon {{\gamma }^{i}})}_{\alpha }}}{4\,r^7}-
    \frac{i\,g^2\,{{{\dot{a}}}_i}\,
       {{(\epsilon {{\gamma }^{i}})}_{\alpha }}}{4\,r^5}+
    \frac{5\,i\,g^2\,{a_i}\,(r \cdot v)\,
       {{(\epsilon {{\gamma }^{i}})}_{\alpha }}}{4\,r^7}
    \brkeq -\frac{35\,i\,g^2\,{v_i}\,{(r \cdot v)}^2\,
       {{(\epsilon {{\gamma }^{i}})}_{\alpha }}}{8\,r^9}+
    \frac{5\,i\,g^2\,{v_i}\,(r \cdot a)\,
       {{(\epsilon {{\gamma }^{i}})}_{\alpha }}}{4\,r^7} .
\end{align}
\shead{Corrections at ${\cal O}(\theta^2)$:}
\begin{align}
\Delta_\epsilon r^m =
 \and -\frac{15\,i\,g^2\,({\dot{\theta}}\theta )\,
       (\epsilon {{\gamma }^m}\theta )}{32\,r^7}+
    \frac{35\,i\,g^2\,{r_m}\,{v_i}\,
       (\epsilon {{\gamma }^j}\theta )\,
       (\theta {{\gamma }^{ij}}\theta )}{64\,r^9}-
    \frac{105\,i\,g^2\,{r_i}\,{v_m}\,
       (\epsilon {{\gamma }^j}\theta )\,
       (\theta {{\gamma }^{ij}}\theta )}{64\,r^9} \brkeq 
   +\frac{35\,i\,g^2\,{r_i}\,{v_j}\,
       (\epsilon {{\gamma }^m}\theta )\,
       (\theta {{\gamma }^{ij}}\theta )}{64\,r^9}+
    \frac{35\,i\,g^2\,(r \cdot v)\,
       (\epsilon {{\gamma }^i}\theta )\,
       (\theta {{\gamma }^{im}}\theta )}{64\,r^9}-
    \frac{5\,i\,g^2\,(\epsilon {{\gamma }^i}{\dot{\theta}})\,
       (\theta {{\gamma }^{im}}\theta )}{32\,r^7} \brkeq 
   -\frac{105\,i\,g^2\,{r_i}\,{v_j}\,
       (\epsilon {{\gamma }^j}\theta )\,
       (\theta {{\gamma }^{im}}\theta )}{64\,r^9}+
    \frac{35\,i\,g^2\,{r_i}\,{v_j}\,
       (\epsilon {{\gamma }^i}\theta )\,
       (\theta {{\gamma }^{jm}}\theta )}{64\,r^9}-
    \frac{105\,i\,g^2\,{r_i}\,{v_j}\,(\epsilon \theta )\,
       (\theta {{\gamma }^{ijm}}\theta )}{64\,r^9} \brkeq 
   +\frac{15\,i\,g^2\,(\epsilon \theta )\,
       ({\dot{\theta}}{{\gamma }^m}\theta )}{32\,r^7}-
    \frac{5\,i\,g^2\,(\epsilon {{\gamma }^i}\theta )\,
       ({\dot{\theta}}{{\gamma }^{im}}\theta )}{32\,r^7} .
\end{align}

\begin{align}
\Delta_\epsilon\theta_\alpha = 
 \and -\frac{175\,i\,g^2\,{r_i}\,{v_j}\,{v_k}\,
       (\theta {{\gamma }^{ik}}\theta )\,
       {{({{\gamma }^{j}}\epsilon )}_{\alpha }}}{64\,r^9}-
    \frac{5\,i\,g^2\,{v_i}\,
       ({\dot{\theta}}{{\gamma }^{ij}}\theta )\,
       {{({{\gamma }^{j}}\epsilon )}_{\alpha }}}{8\,r^7}+
    \frac{35\,i\,g^2\,{r_i}\,(r \cdot v)\,
       ({\dot{\theta}}{{\gamma }^{ij}}\theta )\,
       {{({{\gamma }^{j}}\epsilon )}_{\alpha }}}{16\,r^9}
    \brkeq -\frac{5\,i\,g^2\,{r_i}\,
       ({\dot{\theta}}{{\gamma }^{ij}}{\dot{\theta}})\,
       {{({{\gamma }^{j}}\epsilon )}_{\alpha }}}{32\,r^7}-
    \frac{5\,i\,g^2\,{r_i}\,
       ({\ddot{\theta}}{{\gamma }^{ij}}\theta )\,
       {{({{\gamma }^{j}}\epsilon )}_{\alpha }}}{16\,r^7}-
    \frac{15\,i\,g^2\,{v_i}\,
       ({\dot{\theta}}{{\gamma }^j}\theta )\,
       {{({{\gamma }^{ij}}\epsilon )}_{\alpha }}}{32\,r^7}
    \brkeq +\frac{15\,i\,g^2\,{v_i}\,
       ({\dot{\theta}}{{\gamma }^{ijk}}\theta )\,
       {{({{\gamma }^{jk}}\epsilon )}_{\alpha }}}{64\,r^7}-
    \frac{35\,i\,g^2\,{r_i}\,(r \cdot v)\,
       ({\dot{\theta}}{{\gamma }^{ijk}}\theta )\,
       {{({{\gamma }^{jk}}\epsilon )}_{\alpha }}}{32\,r^9}+
    \frac{5\,i\,g^2\,{r_i}\,
       ({\dot{\theta}}{{\gamma }^{ijk}}{\dot{\theta}})\,
       {{({{\gamma }^{jk}}\epsilon )}_{\alpha }}}{64\,r^7}
    \brkeq +\frac{5\,i\,g^2\,{r_i}\,
       ({\ddot{\theta}}{{\gamma }^{ijk}}\theta )\,
       {{({{\gamma }^{jk}}\epsilon )}_{\alpha }}}{32\,r^7}+
    \frac{105\,i\,g^2\,{r_i}\,{v_j}\,{v_k}\,
       (\theta {{\gamma }^{ikl}}\theta )\,
       {{({{\gamma }^{jl}}\epsilon )}_{\alpha }}}{64\,r^9}+
    \frac{105\,i\,g^2\,v^2\,{r_i}\,
       (\epsilon {{\gamma }^i}\theta )\,{{\theta }_{\alpha }}
       }{32\,r^9} \brkeq 
   -\frac{105\,i\,g^2\,{a_i}\,
       (\epsilon {{\gamma }^i}\theta )\,{{\theta }_{\alpha }}
       }{32\,r^7}+\frac{245\,i\,g^2\,{v_i}\,(r \cdot v)\,
       (\epsilon {{\gamma }^i}\theta )\,{{\theta }_{\alpha }}
       }{16\,r^9}-\frac{315\,i\,g^2\,{r_i}\,{(r \cdot v)}^2\,
       (\epsilon {{\gamma }^i}\theta )\,{{\theta }_{\alpha }}
       }{8\,r^{11}} \brkeq 
   +\frac{35\,i\,g^2\,{r_i}\,(r \cdot a)\,
       (\epsilon {{\gamma }^i}\theta )\,{{\theta }_{\alpha }}
       }{4\,r^9}-\frac{15\,i\,g^2\,{v_i}\,
       (\epsilon {{\gamma }^i}{\dot{\theta}})\,
       {{\theta }_{\alpha }}}{32\,r^7}+
    \frac{35\,i\,g^2\,{r_i}\,(r \cdot v)\,
       (\epsilon {{\gamma }^i}{\dot{\theta}})\,
       {{\theta }_{\alpha }}}{16\,r^9} \brkeq 
   -\frac{5\,i\,g^2\,{r_i}\,
       (\epsilon {{\gamma }^i}{\ddot{\theta}})\,
       {{\theta }_{\alpha }}}{16\,r^7}+
    \frac{105\,i\,g^2\,v^2\,{r_i}\,(\epsilon \theta )\,
       {{({{\gamma }^{i}}\theta )}_{\alpha }}}{16\,r^9}+
    \frac{25\,i\,g^2\,{a_i}\,(\epsilon \theta )\,
       {{({{\gamma }^{i}}\theta )}_{\alpha }}}{32\,r^7}
    \brkeq -\frac{175\,i\,g^2\,{v_i}\,(r \cdot v)\,
       (\epsilon \theta )\,
       {{({{\gamma }^{i}}\theta )}_{\alpha }}}{32\,r^9}+
    \frac{15\,i\,g^2\,{v_i}\,(\epsilon {\dot{\theta}})\,
       {{({{\gamma }^{i}}\theta )}_{\alpha }}}{32\,r^7}-
    \frac{35\,i\,g^2\,{r_i}\,(r \cdot v)\,
       (\epsilon {\dot{\theta}})\,
       {{({{\gamma }^{i}}\theta )}_{\alpha }}}{16\,r^9}
    \brkeq +\frac{5\,i\,g^2\,{r_i}\,
       (\epsilon {\ddot{\theta}})\,
       {{({{\gamma }^{i}}\theta )}_{\alpha }}}{16\,r^7}-
    \frac{105\,i\,g^2\,v^2\,{r_i}\,
       (\epsilon {{\gamma }^{ij}}\theta )\,
       {{({{\gamma }^{j}}\theta )}_{\alpha }}}{64\,r^9}-
    \frac{5\,i\,g^2\,{a_i}\,
       (\epsilon {{\gamma }^{ij}}\theta )\,
       {{({{\gamma }^{j}}\theta )}_{\alpha }}}{32\,r^7}
    \brkeq +\frac{35\,i\,g^2\,{v_i}\,(r \cdot v)\,
       (\epsilon {{\gamma }^{ij}}\theta )\,
       {{({{\gamma }^{j}}\theta )}_{\alpha }}}{32\,r^9}+
    \frac{105\,i\,g^2\,v^2\,{r_i}\,
       (\epsilon {{\gamma }^j}\theta )\,
       {{({{\gamma }^{ij}}\theta )}_{\alpha }}}{64\,r^9}+
    \frac{5\,i\,g^2\,{a_i}\,(\epsilon {{\gamma }^j}\theta )\,
       {{({{\gamma }^{ij}}\theta )}_{\alpha }}}{16\,r^7}
    \brkeq -\frac{35\,i\,g^2\,{v_i}\,(r \cdot v)\,
       (\epsilon {{\gamma }^j}\theta )\,
       {{({{\gamma }^{ij}}\theta )}_{\alpha }}}{16\,r^9}-
    \frac{5\,i\,g^2\,{v_i}\,
       (\epsilon {{\gamma }^j}{\dot{\theta}})\,
       {{({{\gamma }^{ij}}\theta )}_{\alpha }}}{32\,r^7}+
    \frac{35\,i\,g^2\,{r_i}\,(r \cdot v)\,
       (\epsilon {{\gamma }^j}{\dot{\theta}})\,
       {{({{\gamma }^{ij}}\theta )}_{\alpha }}}{16\,r^9}
    \brkeq -\frac{5\,i\,g^2\,{r_i}\,
       (\epsilon {{\gamma }^j}{\ddot{\theta}})\,
       {{({{\gamma }^{ij}}\theta )}_{\alpha }}}{16\,r^7}-
    \frac{105\,i\,g^2\,{r_i}\,{v_j}\,{v_k}\,
       (\epsilon {{\gamma }^j}\theta )\,
       {{({{\gamma }^{ik}}\theta )}_{\alpha }}}{32\,r^9}-
    \frac{5\,i\,g^2\,{v_i}\,(\epsilon {{\gamma }^i}\theta )\,
       {{{\dot{\theta}}}_{\alpha }}}{4\,r^7} \brkeq 
   +\frac{35\,i\,g^2\,{r_i}\,(r \cdot v)\,
       (\epsilon {{\gamma }^i}\theta )\,
       {{{\dot{\theta}}}_{\alpha }}}{16\,r^9}-
    \frac{5\,i\,g^2\,{r_i}\,
       (\epsilon {{\gamma }^i}{\dot{\theta}})\,
       {{{\dot{\theta}}}_{\alpha }}}{16\,r^7}+
    \frac{5\,i\,g^2\,{v_i}\,(\epsilon \theta )\,
       {{({{\gamma }^{i}}{\dot{\theta}})}_{\alpha }}}{4\,r^7}
    \brkeq -\frac{35\,i\,g^2\,{r_i}\,(r \cdot v)\,
       (\epsilon \theta )\,
       {{({{\gamma }^{i}}{\dot{\theta}})}_{\alpha }}}{16\,
       r^9}+\frac{5\,i\,g^2\,{r_i}\,
       (\epsilon {\dot{\theta}})\,
       {{({{\gamma }^{i}}{\dot{\theta}})}_{\alpha }}}{16\,
       r^7}-\frac{5\,i\,g^2\,{v_i}\,
       (\epsilon {{\gamma }^j}\theta )\,
       {{({{\gamma }^{ij}}{\dot{\theta}})}_{\alpha }}}{16\,
       r^7} \brkeq +\frac{35\,i\,g^2\,{r_i}\,(r \cdot v)\,
       (\epsilon {{\gamma }^j}\theta )\,
       {{({{\gamma }^{ij}}{\dot{\theta}})}_{\alpha }}}{16\,
       r^9}-\frac{5\,i\,g^2\,{r_i}\,
       (\epsilon {{\gamma }^j}{\dot{\theta}})\,
       {{({{\gamma }^{ij}}{\dot{\theta}})}_{\alpha }}}{16\,
       r^7}-\frac{5\,i\,g^2\,{r_i}\,
       (\epsilon {{\gamma }^i}\theta )\,
       {{{\ddot{\theta}}}_{\alpha }}}{16\,r^7} \brkeq 
   +\frac{5\,i\,g^2\,{r_i}\,(\epsilon \theta )\,
       {{({{\gamma }^{i}}{\ddot{\theta}})}_{\alpha }}}{16\,
       r^7}-\frac{5\,i\,g^2\,{r_i}\,
       (\epsilon {{\gamma }^j}\theta )\,
       {{({{\gamma }^{ij}}{\ddot{\theta}})}_{\alpha }}}{16\,
       r^7} .
\end{align}
\shead{Corrections at ${\cal O}(\theta^4)$:}
\begin{align}
\Delta_\epsilon r^m =
 \and -\frac{63\,i\,g^2\,{r_i}\,{r_j}\,
       (\epsilon {{\gamma }^k}\theta )\,
       (\theta {{\gamma }^{im}}\theta )\,
       (\theta {{\gamma }^{jk}}\theta )}{128\,r^{11}}+
    \frac{7\,i\,g^2\,(\epsilon {{\gamma }^i}\theta )\,
       (\theta {{\gamma }^{ij}}\theta )\,
       (\theta {{\gamma }^{jm}}\theta )}{128\,r^9} \brkeq 
   -\frac{63\,i\,g^2\,{r_i}\,{r_j}\,
       (\epsilon {{\gamma }^k}\theta )\,
       (\theta {{\gamma }^{ilm}}\theta )\,
       (\theta {{\gamma }^{jkl}}\theta )}{128\,r^{11}}-
    \frac{63\,i\,g^2\,{r_i}\,{r_j}\,(\epsilon \theta )\,
       (\theta {{\gamma }^{ik}}\theta )\,
       (\theta {{\gamma }^{jkm}}\theta )}{128\,r^{11}} .
\end{align}

\begin{align}
\Delta_\epsilon\theta_\alpha = 
 \and -\frac{63\,i\,g^2\,{r_i}\,{r_j}\,{v_k}\,
       (\theta {{\gamma }^{jl}}\theta )\,
       (\theta {{\gamma }^{kl}}\theta )\,
       {{({{\gamma }^{i}}\epsilon )}_{\alpha }}}{256\,r^{11}}
     -\frac{7\,i\,g^2\,{r_i}\,({\dot{\theta}}\theta )\,
       (\theta {{\gamma }^{ij}}\theta )\,
       {{({{\gamma }^{j}}\epsilon )}_{\alpha }}}{128\,r^9}
    \brkeq -\frac{7\,i\,g^2\,{v_i}\,
       (\theta {{\gamma }^{ik}}\theta )\,
       (\theta {{\gamma }^{jk}}\theta )\,
       {{({{\gamma }^{j}}\epsilon )}_{\alpha }}}{256\,r^9}+
    \frac{63\,i\,g^2\,{r_i}\,(r \cdot v)\,
       (\theta {{\gamma }^{ik}}\theta )\,
       (\theta {{\gamma }^{jk}}\theta )\,
       {{({{\gamma }^{j}}\epsilon )}_{\alpha }}}{128\,r^{11}}
    \brkeq +\frac{7\,i\,g^2\,{r_i}\,
       (\theta {{\gamma }^{ijk}}\theta )\,
       ({\dot{\theta}}{{\gamma }^k}\theta )\,
       {{({{\gamma }^{j}}\epsilon )}_{\alpha }}}{32\,r^9}-
    \frac{21\,i\,g^2\,{r_i}\,
       (\theta {{\gamma }^{jk}}\theta )\,
       ({\dot{\theta}}{{\gamma }^{ik}}\theta )\,
       {{({{\gamma }^{j}}\epsilon )}_{\alpha }}}{128\,r^9}
    \brkeq -\frac{7\,i\,g^2\,{r_i}\,
       (\theta {{\gamma }^{ik}}\theta )\,
       ({\dot{\theta}}{{\gamma }^{jk}}\theta )\,
       {{({{\gamma }^{j}}\epsilon )}_{\alpha }}}{64\,r^9}+
    \frac{189\,i\,g^2\,{r_i}\,{r_j}\,{v_k}\,
       (\theta {{\gamma }^{il}}\theta )\,
       (\theta {{\gamma }^{jl}}\theta )\,
       {{({{\gamma }^{k}}\epsilon )}_{\alpha }}}{256\,r^{11}}
    \brkeq +\frac{315\,i\,g^2\,{r_i}\,{r_j}\,{v_k}\,
       (\theta {{\gamma }^{il}}\theta )\,
       (\theta {{\gamma }^{jk}}\theta )\,
       {{({{\gamma }^{l}}\epsilon )}_{\alpha }}}{256\,r^{11}}
     -\frac{189\,i\,g^2\,{r_i}\,{r_j}\,{v_k}\,
       (\theta {{\gamma }^{ilm}}\theta )\,
       (\theta {{\gamma }^{jkm}}\theta )\,
       {{({{\gamma }^{l}}\epsilon )}_{\alpha }}}{256\,r^{11}}
    \brkeq +\frac{7\,i\,g^2\,{r_i}\,
       (\theta {{\gamma }^{jk}}\theta )\,
       ({\dot{\theta}}{{\gamma }^k}\theta )\,
       {{({{\gamma }^{ij}}\epsilon )}_{\alpha }}}{128\,r^9}-
    \frac{63\,i\,g^2\,{r_i}\,{r_j}\,{v_k}\,
       (\theta {{\gamma }^{km}}\theta )\,
       (\theta {{\gamma }^{jlm}}\theta )\,
       {{({{\gamma }^{il}}\epsilon )}_{\alpha }}}{256\,
       r^{11}} \brkeq -\frac{7\,i\,g^2\,{r_i}\,
       ({\dot{\theta}}\theta )\,
       (\theta {{\gamma }^{ijk}}\theta )\,
       {{({{\gamma }^{jk}}\epsilon )}_{\alpha }}}{32\,r^9}-
    \frac{21\,i\,g^2\,{v_i}\,
       (\theta {{\gamma }^{jl}}\theta )\,
       (\theta {{\gamma }^{ikl}}\theta )\,
       {{({{\gamma }^{jk}}\epsilon )}_{\alpha }}}{256\,r^9}
    \brkeq +\frac{63\,i\,g^2\,{r_i}\,(r \cdot v)\,
       (\theta {{\gamma }^{jl}}\theta )\,
       (\theta {{\gamma }^{ikl}}\theta )\,
       {{({{\gamma }^{jk}}\epsilon )}_{\alpha }}}{128\,
       r^{11}}+\frac{7\,i\,g^2\,{r_i}\,
       (\theta {{\gamma }^{jk}}\theta )\,
       ({\dot{\theta}}{{\gamma }^i}\theta )\,
       {{({{\gamma }^{jk}}\epsilon )}_{\alpha }}}{128\,r^9}
    \brkeq +\frac{35\,i\,g^2\,{r_i}\,
       (\theta {{\gamma }^{ij}}\theta )\,
       ({\dot{\theta}}{{\gamma }^k}\theta )\,
       {{({{\gamma }^{jk}}\epsilon )}_{\alpha }}}{128\,r^9}+
    \frac{7\,i\,g^2\,{r_i}\,
       (\theta {{\gamma }^{ijl}}\theta )\,
       ({\dot{\theta}}{{\gamma }^{kl}}\theta )\,
       {{({{\gamma }^{jk}}\epsilon )}_{\alpha }}}{64\,r^9}
    \brkeq -\frac{21\,i\,g^2\,{r_i}\,
       (\theta {{\gamma }^{jl}}\theta )\,
       ({\dot{\theta}}{{\gamma }^{ikl}}\theta )\,
       {{({{\gamma }^{jk}}\epsilon )}_{\alpha }}}{128\,r^9}+
    \frac{189\,i\,g^2\,{r_i}\,{r_j}\,{v_k}\,
       (\theta {{\gamma }^{im}}\theta )\,
       (\theta {{\gamma }^{jlm}}\theta )\,
       {{({{\gamma }^{kl}}\epsilon )}_{\alpha }}}{256\,
       r^{11}} \brkeq -\frac{63\,i\,g^2\,{r_i}\,{r_j}\,
       {v_k}\,(\theta {{\gamma }^{il}}\theta )\,
       (\theta {{\gamma }^{jkm}}\theta )\,
       {{({{\gamma }^{lm}}\epsilon )}_{\alpha }}}{64\,r^{11}}
     +\frac{63\,i\,g^2\,{r_i}\,{r_j}\,{v_k}\,
       (\theta {{\gamma }^{ik}}\theta )\,
       (\theta {{\gamma }^{jlm}}\theta )\,
       {{({{\gamma }^{lm}}\epsilon )}_{\alpha }}}{256\,
       r^{11}} \brkeq 
+\frac{35\,i\,g^2\,{r_i}\,
       ({\dot{\theta}}\theta )\,
       (\epsilon {{\gamma }^i}\theta )\,{{\theta }_{\alpha }}
       }{64\,r^9}-\frac{7\,i\,g^2\,{v_i}\,
       (\epsilon {{\gamma }^j}\theta )\,
       (\theta {{\gamma }^{ij}}\theta )\,
       {{\theta }_{\alpha }}}{128\,r^9} \brkeq 
   -\frac{63\,i\,g^2\,{r_i}\,(r \cdot v)\,
       (\epsilon {{\gamma }^j}\theta )\,
       (\theta {{\gamma }^{ij}}\theta )\,
       {{\theta }_{\alpha }}}{128\,r^{11}}-
    \frac{35\,i\,g^2\,{r_i}\,(\epsilon \theta )\,
       ({\dot{\theta}}{{\gamma }^i}\theta )\,
       {{\theta }_{\alpha }}}{64\,r^9} \nonumber
\end{align}
\begin{align}
&   +\frac{35\,i\,g^2\,{r_i}\,
       (\epsilon {{\gamma }^j}\theta )\,
       ({\dot{\theta}}{{\gamma }^{ij}}\theta )\,
       {{\theta }_{\alpha }}}{64\,r^9}+
    \frac{35\,i\,g^2\,{r_i}\,(\epsilon \theta )\,
       ({\dot{\theta}}\theta )\,
       {{({{\gamma }^{i}}\theta )}_{\alpha }}}{64\,r^9}
    \brkeq -\frac{189\,i\,g^2\,{r_i}\,{r_j}\,{v_k}\,
       (\epsilon \theta )\,(\theta {{\gamma }^{jk}}\theta )\,
       {{({{\gamma }^{i}}\theta )}_{\alpha }}}{64\,r^{11}}+
    \frac{63\,i\,g^2\,{r_i}\,{r_j}\,{v_k}\,
       (\epsilon {{\gamma }^l}\theta )\,
       (\theta {{\gamma }^{jkl}}\theta )\,
       {{({{\gamma }^{i}}\theta )}_{\alpha }}}{32\,r^{11}}
    \brkeq -\frac{35\,i\,g^2\,{r_i}\,
       (\epsilon {{\gamma }^j}\theta )\,
       ({\dot{\theta}}{{\gamma }^j}\theta )\,
       {{({{\gamma }^{i}}\theta )}_{\alpha }}}{64\,r^9}-
    \frac{21\,i\,g^2\,{v_i}\,(\epsilon \theta )\,
       (\theta {{\gamma }^{ij}}\theta )\,
       {{({{\gamma }^{j}}\theta )}_{\alpha }}}{128\,r^9}
    \brkeq +\frac{63\,i\,g^2\,{r_i}\,(r \cdot v)\,
       (\epsilon \theta )\,(\theta {{\gamma }^{ij}}\theta )\,
       {{({{\gamma }^{j}}\theta )}_{\alpha }}}{32\,r^{11}}-
    \frac{7\,i\,g^2\,{r_i}\,(\epsilon {\dot{\theta}})\,
       (\theta {{\gamma }^{ij}}\theta )\,
       {{({{\gamma }^{j}}\theta )}_{\alpha }}}{64\,r^9}
    \brkeq +\frac{7\,i\,g^2\,{r_i}\,
       (\epsilon {{\gamma }^j}\theta )\,
       ({\dot{\theta}}{{\gamma }^i}\theta )\,
       {{({{\gamma }^{j}}\theta )}_{\alpha }}}{64\,r^9}-
    \frac{7\,i\,g^2\,{r_i}\,(\epsilon {{\gamma }^i}\theta )\,
       ({\dot{\theta}}{{\gamma }^j}\theta )\,
       {{({{\gamma }^{j}}\theta )}_{\alpha }}}{64\,r^9}
    \brkeq -\frac{35\,i\,g^2\,{r_i}\,(\epsilon \theta )\,
       ({\dot{\theta}}{{\gamma }^{ij}}\theta )\,
       {{({{\gamma }^{j}}\theta )}_{\alpha }}}{64\,r^9}-
    \frac{35\,i\,g^2\,{v_i}\,
       (\epsilon {{\gamma }^j}\theta )\,
       (\theta {{\gamma }^{ijk}}\theta )\,
       {{({{\gamma }^{k}}\theta )}_{\alpha }}}{128\,r^9} 
    \brkeq +\frac{63\,i\,g^2\,{r_i}\,(r \cdot v)\,
       (\epsilon {{\gamma }^j}\theta )\,
       (\theta {{\gamma }^{ijk}}\theta )\,
       {{({{\gamma }^{k}}\theta )}_{\alpha }}}{32\,r^{11}}-
    \frac{7\,i\,g^2\,{r_i}\,
       (\epsilon {{\gamma }^j}{\dot{\theta}})\,
       (\theta {{\gamma }^{ijk}}\theta )\,
       {{({{\gamma }^{k}}\theta )}_{\alpha }}}{64\,r^9} 
    \brkeq -\frac{35\,i\,g^2\,{r_i}\,
       (\epsilon {{\gamma }^j}\theta )\,
       ({\dot{\theta}}{{\gamma }^{ijk}}\theta )\,
       {{({{\gamma }^{k}}\theta )}_{\alpha }}}{64\,r^9}+
    \frac{7\,i\,g^2\,{r_i}\,({\dot{\theta}}\theta )\,
       (\epsilon {{\gamma }^j}\theta )\,
       {{({{\gamma }^{ij}}\theta )}_{\alpha }}}{16\,r^9}
    \brkeq -\frac{7\,i\,g^2\,{r_i}\,(\epsilon \theta )\,
       ({\dot{\theta}}{{\gamma }^j}\theta )\,
       {{({{\gamma }^{ij}}\theta )}_{\alpha }}}{16\,r^9}-
    \frac{7\,i\,g^2\,{v_i}\,(\epsilon {{\gamma }^j}\theta )\,
       (\theta {{\gamma }^{jk}}\theta )\,
       {{({{\gamma }^{ik}}\theta )}_{\alpha }}}{128\,r^9}
    \brkeq +\frac{7\,i\,g^2\,{r_i}\,
       (\epsilon {{\gamma }^j}{\dot{\theta}})\,
       (\theta {{\gamma }^{jk}}\theta )\,
       {{({{\gamma }^{ik}}\theta )}_{\alpha }}}{64\,r^9}+
    \frac{63\,i\,g^2\,{r_i}\,{r_j}\,{v_k}\,
       (\epsilon {{\gamma }^l}\theta )\,
       (\theta {{\gamma }^{jl}}\theta )\,
       {{({{\gamma }^{ik}}\theta )}_{\alpha }}}{32\,r^{11}}
    \brkeq -\frac{63\,i\,g^2\,{r_i}\,{r_j}\,{v_k}\,
       (\epsilon {{\gamma }^l}\theta )\,
       (\theta {{\gamma }^{jk}}\theta )\,
       {{({{\gamma }^{il}}\theta )}_{\alpha }}}{128\,r^{11}}+
    \frac{189\,i\,g^2\,{r_i}\,{r_j}\,{v_k}\,
       (\epsilon {{\gamma }^k}\theta )\,
       (\theta {{\gamma }^{jl}}\theta )\,
       {{({{\gamma }^{il}}\theta )}_{\alpha }}}{128\,r^{11}}
    \brkeq +\frac{189\,i\,g^2\,{r_i}\,{r_j}\,{v_k}\,
       (\epsilon \theta )\,
       (\theta {{\gamma }^{jkl}}\theta )\,
       {{({{\gamma }^{il}}\theta )}_{\alpha }}}{128\,r^{11}}-
    \frac{63\,i\,g^2\,{r_i}\,{r_j}\,{v_k}\,
       (\epsilon {{\gamma }^i}\theta )\,
       (\theta {{\gamma }^{kl}}\theta )\,
       {{({{\gamma }^{jl}}\theta )}_{\alpha }}}{128\,r^{11}}
    \brkeq -\frac{63\,i\,g^2\,{r_i}\,{r_j}\,{v_k}\,
       (\epsilon \theta )\,(\theta {{\gamma }^{jl}}\theta )\,
       {{({{\gamma }^{ikl}}\theta )}_{\alpha }}}{128\,r^{11}}
     -\frac{63\,i\,g^2\,{r_i}\,{r_j}\,{v_k}\,
       (\epsilon {{\gamma }^l}\theta )\,
       (\theta {{\gamma }^{jlm}}\theta )\,
       {{({{\gamma }^{ikm}}\theta )}_{\alpha }}}{128\,r^{11}}
    \brkeq +\frac{35\,i\,g^2\,{r_i}\,
       (\epsilon {{\gamma }^j}\theta )\,
       (\theta {{\gamma }^{ij}}\theta )\,
       {{{\dot{\theta}}}_{\alpha }}}{64\,r^9}-
    \frac{7\,i\,g^2\,{r_i}\,(\epsilon \theta )\,
       (\theta {{\gamma }^{ij}}\theta )\,
       {{({{\gamma }^{j}}{\dot{\theta}})}_{\alpha }}}{16\,
       r^9} \brkeq -\frac{7\,i\,g^2\,{r_i}\,
       (\epsilon {{\gamma }^j}\theta )\,
       (\theta {{\gamma }^{ijk}}\theta )\,
       {{({{\gamma }^{k}}{\dot{\theta}})}_{\alpha }}}{16\,
       r^9}-\frac{7\,i\,g^2\,{r_i}\,
       (\epsilon {{\gamma }^j}\theta )\,
       (\theta {{\gamma }^{jk}}\theta )\,
       {{({{\gamma }^{ik}}{\dot{\theta}})}_{\alpha }}}{64\,
       r^9} .
\end{align}
\shead{Corrections at ${\cal O}(\theta^6)$:}
\begin{align} 
& \Delta_\epsilon\theta_\alpha = \brkeq
 +\frac{21\,i\,g^2\,{r_i}\,(\theta {{\gamma }^{il}}\theta )\,
       (\theta {{\gamma }^{jk}}\theta )\,
       (\theta {{\gamma }^{kl}}\theta )\,
       {{({{\gamma }^{j}}\epsilon )}_{\alpha }}}{1024\,
       r^{11}}-\frac{231\,i\,g^2\,{r_i}\,{r_j}\,{r_k}\,
       (\theta {{\gamma }^{il}}\theta )\,
       (\theta {{\gamma }^{jm}}\theta )\,
       (\theta {{\gamma }^{km}}\theta )\,
       {{({{\gamma }^{l}}\epsilon )}_{\alpha }}}{1024\,
       r^{13}} \brkeq +\frac{231\,i\,g^2\,{r_i}\,{r_j}\,{r_k}\,
       (\theta {{\gamma }^{im}}\theta )\,
       (\theta {{\gamma }^{jmn}}\theta )\,
       (\theta {{\gamma }^{kln}}\theta )\,
       {{({{\gamma }^{l}}\epsilon )}_{\alpha }}}{1024\,
       r^{13}}+\frac{21\,i\,g^2\,{r_i}\,
       (\theta {{\gamma }^{jl}}\theta )\,
       (\theta {{\gamma }^{lm}}\theta )\,
       (\theta {{\gamma }^{ikm}}\theta )\,
       {{({{\gamma }^{jk}}\epsilon )}_{\alpha }}}{1024\,
       r^{11}} \brkeq +\frac{21\,i\,g^2\,{r_i}\,
       (\theta {{\gamma }^{jl}}\theta )\,
       (\theta {{\gamma }^{km}}\theta )\,
       (\theta {{\gamma }^{ilm}}\theta )\,
       {{({{\gamma }^{jk}}\epsilon )}_{\alpha }}}{2048\,
       r^{11}}-\frac{231\,i\,g^2\,{r_i}\,{r_j}\,{r_k}\,
       (\theta {{\gamma }^{il}}\theta )\,
       (\theta {{\gamma }^{jn}}\theta )\,
       (\theta {{\gamma }^{kmn}}\theta )\,
       {{({{\gamma }^{lm}}\epsilon )}_{\alpha }}}{1024\,
       r^{13}} \brkeq -\frac{231\,i\,g^2\,{r_i}\,{r_j}\,{r_k}\,
       (\theta {{\gamma }^{iln}}\theta )\,
       (\theta {{\gamma }^{jnq}}\theta )\,
       (\theta {{\gamma }^{kmq}}\theta )\,
       {{({{\gamma }^{lm}}\epsilon )}_{\alpha }}}{2048\,
       r^{13}}+\frac{21\,i\,g^2\,{r_i}\,(\epsilon \theta )\,
       (\theta {{\gamma }^{ik}}\theta )\,
       (\theta {{\gamma }^{jk}}\theta )\,
       {{({{\gamma }^{j}}\theta )}_{\alpha }}}{512\,r^{11}}
    \brkeq +\frac{231\,i\,g^2\,{r_i}\,{r_j}\,{r_k}\,
       (\epsilon \theta )\,(\theta {{\gamma }^{il}}\theta )\,
       (\theta {{\gamma }^{jl}}\theta )\,
       {{({{\gamma }^{k}}\theta )}_{\alpha }}}{512\,r^{13}}+
    \frac{21\,i\,g^2\,{r_i}\,(\epsilon {{\gamma }^j}\theta )\,
       (\theta {{\gamma }^{kl}}\theta )\,
       (\theta {{\gamma }^{ijl}}\theta )\,
       {{({{\gamma }^{k}}\theta )}_{\alpha }}}{512\,r^{11}}
    \brkeq +\frac{21\,i\,g^2\,{r_i}\,
       (\epsilon {{\gamma }^j}\theta )\,
       (\theta {{\gamma }^{jl}}\theta )\,
       (\theta {{\gamma }^{ikl}}\theta )\,
       {{({{\gamma }^{k}}\theta )}_{\alpha }}}{512\,r^{11}}+
    \frac{231\,i\,g^2\,{r_i}\,{r_j}\,{r_k}\,
       (\epsilon {{\gamma }^l}\theta )\,
       (\theta {{\gamma }^{im}}\theta )\,
       (\theta {{\gamma }^{jlm}}\theta )\,
       {{({{\gamma }^{k}}\theta )}_{\alpha }}}{512\,r^{13}}
    \brkeq -\frac{21\,i\,g^2\,{r_i}\,
       (\epsilon {{\gamma }^j}\theta )\,
       (\theta {{\gamma }^{jl}}\theta )\,
       (\theta {{\gamma }^{kl}}\theta )\,
       {{({{\gamma }^{ik}}\theta )}_{\alpha }}}{512\,r^{11}}+
    \frac{231\,i\,g^2\,{r_i}\,{r_j}\,{r_k}\,(\epsilon \theta )\,
       (\theta {{\gamma }^{im}}\theta )\,
       (\theta {{\gamma }^{jlm}}\theta )\,
       {{({{\gamma }^{kl}}\theta )}_{\alpha }}}{512\,r^{13}}
    \brkeq -\frac{231\,i\,g^2\,{r_i}\,{r_j}\,{r_k}\,
       (\epsilon {{\gamma }^l}\theta )\,
       (\theta {{\gamma }^{il}}\theta )\,
       (\theta {{\gamma }^{jm}}\theta )\,
       {{({{\gamma }^{km}}\theta )}_{\alpha }}}{512\,r^{13}}+
    \frac{231\,i\,g^2\,{r_i}\,{r_j}\,{r_k}\,
       (\epsilon {{\gamma }^l}\theta )\,
       (\theta {{\gamma }^{iln}}\theta )\,
       (\theta {{\gamma }^{jmn}}\theta )\,
       {{({{\gamma }^{km}}\theta )}_{\alpha }}}{512\,r^{13}}.
\end{align}

\end{document}